%% file: main.tex
\documentclass[sigconf]{acmart}
\AtBeginDocument{%
  }

\usepackage[capitalize, noabbrev]{cleveref}
\usepackage[table]{xcolor}
\usepackage{fontawesome} 
\usepackage{longtable}
\usepackage{tabularx}
\usepackage{multirow}
\usepackage{listings}
\usepackage{siunitx}

\copyrightyear{2026}
\acmYear{2026}
\setcopyright{cc}
\setcctype{by}
\acmConference[CHI '26]{Proceedings of the 2026 CHI Conference on Human Factors in Computing Systems}{April 13--17, 2026}{Barcelona, Spain}
\acmBooktitle{Proceedings of the 2026 CHI Conference on Human Factors in Computing Systems (CHI '26), April 13--17, 2026, Barcelona, Spain}
\acmPrice{}
\acmDOI{10.1145/3772318.3790648}
\acmISBN{979-8-4007-2278-3/2026/04}

\definecolor{colortodo}{rgb}{0.9, 0.6, 0}    
\definecolor{colordaniel}{rgb}{0, 0.7, 0.35}
\definecolor{colorflo}{rgb}{0.2, 0.4, 0.8}
\definecolor{colorkrystsina}{rgb}{0.8, 0.3, 0.5} 
\definecolor{colorrevision}{rgb}{0, 0.2, 0.8}
\definecolor{colorrevisionCR}{rgb}{0, 0.8, 0.2}

\definecolor{colorTableGray}{gray}{0.95}

\newcommand{\revision}[1]{#1}
\newcommand{\revisionCR}[1]{#1}

\newcommand{\p}[2]{$P_{#1,#2}$}

\newcommand{\groupsN}{14}
\newcommand{\studyN}{30}

\definecolor{colorquote}{rgb}{0.45, 0.45, 0.45}  
\newcommand{\pquote}[1]{``#1''} %

\newcommand{\emphresult}[1]{\textbf{#1}}

\begin{document}

\title{Collaborative Document Editing with Multiple Users and AI Agents}

\author{Florian Lehmann}
\authornote{Both authors contributed equally to this research.}
\orcid{0000-0003-0201-867X}
\email{florian.lehmann@uni-bayreuth.de}
\affiliation{%
  \institution{University of Bayreuth}
  \streetaddress{Universit\"atsstra{\ss}e 30}
  \city{Bayreuth}
  \country{Germany}
  \postcode{95447}
}

\author{Krystsina Shauchenka}
\authornotemark[1]
\orcid{0009-0005-9175-7248}
\email{krystsina.shauchenka@uni-bayreuth.de}
\affiliation{%
  \institution{University of Bayreuth}
  \streetaddress{Universit\"atsstra{\ss}e 30}
  \city{Bayreuth}
  \country{Germany}
  \postcode{95447}
}

\author{Daniel Buschek}
\orcid{0000-0002-0013-715X}
\email{daniel.buschek@uni-bayreuth.de}
\affiliation{%
  \institution{University of Bayreuth}
  \streetaddress{Universit\"atsstra{\ss}e 30}
  \city{Bayreuth}
  \country{Germany}
  \postcode{95447}
}

\renewcommand{\shortauthors}{Lehmann et al.}

\begin{abstract}

\input{sections/abstract}

\end{abstract}


\begin{teaserfigure}
   \includegraphics[width=\textwidth]{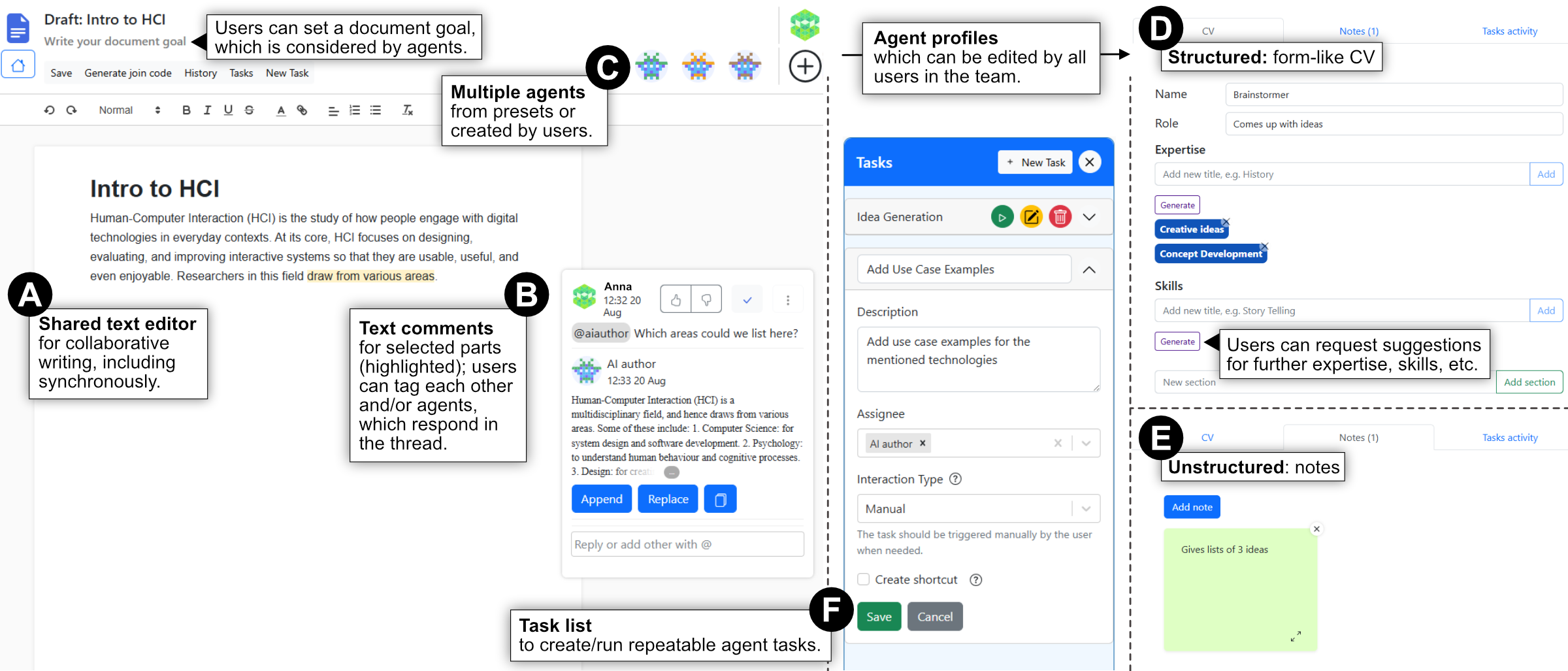}
   \caption{Overview of our prototype: A shared document editor (A) with comments (B), which co-writers can also use to delegate tasks to AI agents (C). Users create custom agents through profiles, via a structured form-like ``CV'' (D) and/or unstructured notes (E). Agents can be assigned to repeatable tasks in a task list (F). Note: This task list (F) opens as a sidebar that replaces the comments sidebar (B), while (D) and (E) are part of a separate full-screen profile view, shown here in parallel for illustration.}
   \Description{The figure presents an overview of the prototype system. On the left, a shared text editor is shown with highlighted passages and threaded comments for collaboration. At the top, users can set document goals and choose from multiple AI agents. In the centre-right, a sidebar displays a task list where repeatable tasks can be assigned to agents. On the far right, an agent profile view is illustrated, including a structured form-like CV with expertise and skills, and an unstructured notes area for free-form information.}
   \label{fig:teaser}
 \end{teaserfigure}

\maketitle

\input{sections/introduction}
\input{sections/related_work}

\input{sections/concept}
\input{sections/system}
\input{sections/user_study}
\input{sections/results}

\input{sections/discussion}

\input{sections/conclusion}

\begin{acks}
This project is funded by the Deutsche Forschungsgemeinschaft (DFG, German Research Foundation) -- 525037874. During parts of this project, Daniel Buschek was supported by a Google Research Scholar Award.
\end{acks}

\bibliographystyle{ACM-Reference-Format}
\bibliography{bibliography}

\appendix
\input{sections/appendix}

\end{document}

%% file: sections/abstract.tex
Current AI writing support tools are largely designed for individuals, complicating collaboration when co-writers must leave the shared workspace to use AI and then communicate and reintegrate results. We propose integrating AI agents directly into collaborative writing environments. Our prototype makes AI use \revision{visible to all users} through two new shared objects: \revision{user-defined} agent profiles and tasks. Agent responses appear in the familiar comment feature. In a user study (N=\studyN), \groupsN{} teams worked on writing projects during one week. Interaction logs and interviews show that teams incorporated agents into existing norms of authorship, control, and coordination, rather than treating them as team members. Agent profiles were viewed as personal territory, while created agents and outputs became shared resources. We discuss implications for team-based AI interaction, highlighting opportunities and boundaries for treating AI as a shared resource in collaborative work.

%% file: sections/introduction.tex
\section{Introduction}

People write \textit{with} people. Teams co-author notes, reports, and academic papers like this one, and even a single-author novel evolves through (written) comments from editors and early readers.
Increasingly, teams also write ``with AI'' -- or do they?

Today, co-writers can get AI support, such as with recent writing tools and chatbots based on Large Language Models (LLMs)~\cite{Lee2024designspace}.
However, \revision{since} these tools are currently designed for individual use, %
co-writers must leave the shared workspace to interact with AI and later remember, integrate, and communicate the results with their collaborators. \revision{Not limited to AI, such tool switches away from collaborative documents are known to be challenging~\cite{LarsenLedet2020collabwriting}.}

In this paper, we explore a new design direction: direct integration of customisable, shared AI agents into collaborative documents. \revision{With ``customisable'', we concretely mean allowing users to define their own agents and tasks for those.}

Moving beyond the design focus on a single writer's use of AI is both timely and important: From a usability perspective, AI should support diverse workflows, including the complex dynamics of collaborative writing. Pragmatically, there is growing industry demand to integrate new AI features into widely used tools that already offer real-time collaboration. Finally, through a human-centred lens, if HCI is to shape AI that augments rather than replaces human collaboration, we need designs that enhance teamwork rather than push team members out of their shared context.

Integrating AI into collaborative documents presents a challenging design problem: Enabling teams to use AI together demands addressing multiple aspects at the same time, such as collaborative prompting~\cite{Han2024teamsai} and task delegation~\cite{Tankelevitch2024metacog}, maintaining transparency, and respecting individual and shared needs and preferences~\cite{LarsenLedet2021dontagree}. Prompting AI effectively is already difficult for individuals~\cite{ZamfirescuPereira2023johnny} and this challenge is amplified in team contexts, where goals and perspectives must be negotiated \citep{Han2024teamsai}. Moreover, there are currently no established UI design patterns for AI in collaborative writing. This is reflected in recent surveys and design theory that focus on individual writing~\cite{Buschek2024collage, Lee2024designspace, Reza2025cowritingai}. As recently concluded by \citet{Chen2025nonnativecollab}, ``we have not yet seen AI-assisted writing tools targeting collaborative writing.''

More broadly, existing research has explored collaborative writing among humans~\cite{Birnholtz2012tracking, Birnholtz2013writeherenow, Bodker2024material, LarsenLedet2019territory, LarsenLedet2020collabwriting, LarsenLedet2021dontagree}, interfaces for individual writing with AI~\cite{Lee2024designspace}, or crowd work contexts~\cite{Bernstein2010soylent, Feldman2021crowds, Huang2020heteroglossia}, where the writer in focus can also involve AI~\cite{Huang2025inspo}. Related, AI-mediated communication (AIMC)~\cite{Fu2024texttoself, Hancock2020aimc, Liu2022aimail} typically involves AI and written messages between people but not teams writing on shared documents. Beyond writing, team use of AI has been explored for group ideation~\cite{Han2024teamsai, Shaer2024aibrainwriting, Shin2023collabIdeationWorkshop} and team organisation~\cite{Asthana2025meetingrecap, Morrison2024aireminders}. 

In sum, research has focused on collaborative writing or on writing with AI, but not on the intersection. This motivates our research questions \revision{on collaborative use, specifically on how teams experience and work with agents integrated into collaborative writing}:
\begin{itemize}
    \item \textbf{RQ1:} How do co-writers \textit{create} AI agents in a collaborative setting?
    \item \textbf{RQ2:} How do co-writers \revision{collaboratively} \textit{interact with} shared AI agents?
\end{itemize}

To address these questions, we developed a functional prototype of a collaborative text editor that embeds AI agents (\cref{fig:teaser}). Our design introduces two novel shared objects: (1) \textit{agent profiles}, allowing teams to configure AI personas, and (2) \textit{agent tasks}, which make AI task delegation explicit and visible to all collaborators. AI responses are integrated via collaborative comments, enabling co-writers to engage with AI output within a familiar UI concept.

We deployed this prototype as a \revision{technology probe~\cite{Feng2025canvil, Hutchinson2003techprobe}}, giving it for a week to \studyN{} participants working in \groupsN{} groups. 
Our analysis of interaction logs and interviews revealed how teams incorporate AI into existing norms of authorship, control, and coordination, rather than treating agents as equal team members. Agent profiles were treated as personal territory, while created agents and their outputs became shared resources.
Teams also deliberated over the number of agents. Some preferred a single, general-purpose agent \revision{because they perceived it as efficient}, while others saw multiple role-specific agents as an investment into future prompting and diverse perspectives.

In summary, this paper contributes (1) a UI and interaction design concept and prototype for shared AI agents in collaborative writing, (2) an empirical study of team-AI interaction behaviour and perception, and (3) design insights for integrating AI into collaborative documents. In a broad view, our findings contribute to the literature on team-AI interaction and writing tools by revealing both opportunities and boundaries for treating AI as a shared resource in collaboration.

%% file: sections/related_work.tex
\section{Background and Related Work}
We discuss related work on collaborative writing, writing tools that involve AI, and group interaction with AI. %

\subsection{Tools for Writing with AI and Others}

First, we review research on writing tools that involve AI and/or other people.

\subsubsection{Intelligent interactive writing tools are predominantly designed for a single user} 
Recently, \citet{Lee2024designspace} surveyed 115 intelligent writing tools. None of these designs aims at writing with \textit{multiple users} plus AI. References to ``collaborative'' writing were limited to (1) paired text suggestions from two language models~\cite{Clark2021choose} and (2) human-AI co-writing with a single user (e.g.~\cite{Babaian2002assistant, Biermann2022companion, Chakrabarty2022poem, Lee2022coauthor}). Similarly, \citet{Ocampo2025beyond} use ``collaborative editor'' to mean human+AI, not multiple people. We found an exception in \textit{SAGA}~\cite{Shakeri2021saga}, where two people and GPT-3 alternated lines in a story. Recent work continues to focus on single-user co-writing, including another review~\cite{Reza2025cowritingai} and work published after the above survey in CSCW~\cite{Hwang2025authenticity, Sun2024metawriter, Wan2024secondmind}, an HCI venue traditionally associated with human collaboration. Consequently, the design space covered so far~\cite{Lee2024designspace} omits multi-user aspects and their integration with AI. This gap motivates our work.

\subsubsection{Writing support from the crowd and AI}
Some research proposed text editors that involve crowd workers~\cite{Bernstein2010soylent, Feldman2021crowds, Huang2020heteroglossia, Huang2025inspo, Kim2014ensemble}, where multiple people (and sometimes AI~\cite{Huang2025inspo}) contribute text. However, these systems still focus on one main writer who requests support rather than team collaboration. For instance, in \textit{Heteroglossia}~\cite{Huang2020heteroglossia} and \textit{Inspo}~\cite{Huang2025inspo}, the writer highlights a span to request suggestions from crowd workers or an LLM, which then appear in a sidebar. That is, the interaction design does not involve multiple writers in jointly managing AI input or evaluating its results. Addressing these gaps motivates our design exploration.

\subsection{Group Interaction with AI}
Here, we review group interaction with AI, not limited to writing.

\subsubsection{Group ideation with AI}
Recent HCI work has examined group ideation with AI, such as the CHI'23 workshop by \citet{Shin2023collabIdeationWorkshop}. \citet{Han2024teamsai} studied synchronous collaboration between two people designing stage sets, where jointly crafting prompts for ChatGPT and Midjourney was both helpful (e.g. finding words, sharing opinions) and costly (e.g. debating prompting strategies). \citet{Shaer2024aibrainwriting} explored group brainwriting with a canvas UI and access to ChatGPT, but without direct integration. 
In contrast, we investigate collaborative AI use in a custom text editor with dedicated features (e.g. comments, AI tasks, agent profiles) that integrate prompting and writing for synchronous and asynchronous collaboration.

\subsubsection{Group organisation and AI-mediated communication}
Another related area explores AI support for team onboarding~\cite{Shin2023introbot} and coordination, such as facilitating discussions~\cite{Kim2020botinthebunch} or generating task reminders~\cite{Morrison2024aireminders} and meeting recaps~\cite{Asthana2025meetingrecap}. Similar features have recently been added to team software.\footnote{E.g. AI-generated meeting summaries in \href{https://support.zoom.com/hc/en/article?id=zm_kb&sysparm_article=KB0080263}{Zoom}, \href{https://techcommunity.microsoft.com/blog/microsoftteamsblog/intelligent-meeting-recap-in-teams-premium-now-available/3832541}{MS Teams}, and \href{https://workspace.google.com/resources/ai-for-meetings/}{Google Meet}.}

More broadly, our work differs from AI-mediated communication (AIMC)~\cite{Hancock2020aimc}. While there is conceptual overlap (text, multiple people, AI), AIMC typically concerns text created by one person to communicate with another (e.g. emails~\cite{Liu2022aimail}), not multiple people co-authoring a shared document with AI. For example, the recent diary study of AIMC tools by \citet{Fu2024texttoself} did not surface such collaborative use.

\subsection{Collaborative Writing and Adding AI}
Next, we reflect on insights into collaborative writing (without AI) and how these motivated the features we explore.

\subsubsection{Bringing AI into the collaborative home of writing}

\revision{The notion of \textit{artifact ecologies} %
deals with how collaborative writers involve multiple tools and transition between them: \citet{LarsenLedet2020collabwriting} found that such transitions introduce challenges, such as ``When writers work outside the common documents and applications, their co-writers lose sense of the progress and who is working on what.''} This maps well to the situation with AI tools today, which are separated from shared documents. \citet{LarsenLedet2020collabwriting} \revision{also} describe the concept of a \textit{collaborative home}, where writing occurs in a synchronized platform \revision{such} that all writers see the same content and tools/UI. \revision{Motivated by these prior observations on transition costs and shared tools}, our design explores how \revision{teams collaborate and use AI integrated} directly into the collaborative home.

\subsubsection{Integrating AI into tools and text as material for collaboration}
\citet{Bodker2024material} examine how material qualities of tools and text support collaboration. For example, co-writers use \textit{comments} to reference others for planning and task delegation (e.g. ``Sue will do this''), linking cooperative activities (e.g. added as a result of a meeting) with coordination (e.g. serves as a reminder to others and Sue herself). Similarly, \textit{written plans} (e.g. outlines, todo lists) ``provide a material coupling between the thinking and the material production of text.''

Building on this, our design integrates AI agents with such materials. In our \textit{comment} feature, users can reference not only each other but also AI agents, which reply in the thread. Users can also create repeatable \textit{plans} as a task list, executable via a button or automatically (e.g. when saving the document).

Further work identified ``writing territory''~\cite{LarsenLedet2019territory} and showed that edits and comments carry social meaning (e.g. edits as criticism)~\cite{Birnholtz2012tracking, Birnholtz2013writeherenow, Park2023whywhy}. To respect these dynamics and keep users in control, we designed agents to never edit the text directly. Instead, they add task results as comments, linking the two materials.

\subsubsection{Supporting idiosyncratic preferences and common objects, also for AI}
Co-writers' processes often differ. In a co-design study with academic writers, \citet{LarsenLedet2021dontagree} found contrasting needs and desires, complicating one-size-fits-all solutions. This motivates our exploration of support for defining multiple AI agents, which co-writers can use to tailor agents to different needs. At the same time, all agents remain visible and usable to everyone, making it possible for teams to treat them  as \textit{common objects}~\cite{LarsenLedet2020collabwriting}, alongside the text itself.

\begin{table*}[t]
    \setlength{\tabcolsep}{3pt} 
    \renewcommand{\arraystretch}{1.35} 
    \newcolumntype{Y}{>{\raggedright\arraybackslash}X}
  \centering
  \caption{\revision{Our main design elements, compared to related work. The overarching novelty is that we allow for \textit{collaboratively} creating and interacting with agents and tasks, \textit{directly within} a shared document editor, including in \textit{comments}.}}
  \Description{Four-column table summarising design elements, their use in related work, our design decisions, and motivations, also in the context of a technology probe study.}
  \label{tab:design_elements}
  \footnotesize
  \begin{tabularx}{\linewidth}{@{}>{\raggedright\arraybackslash}p{1.5cm} Y Y Y Y @{}}
    \toprule
    \textbf{Design element} & \textbf{Examples in related work and products} & \textbf{In our design} & \textbf{Motivation for our design} & \textbf{Motivation in context of our technology probe study} \\
    \midrule
    Multiple agent personas in the context of writing & \citet{Benharrak2024aipersonas}: A single user can define multiple agents with a structured UI and request feedback on text, shown in a sidebar. \citet{Siddiqui2025scriptshift}: A single user can request feedback on text from pre-defined agents, shown in a pop-up. & Multiple users can define multiple agents, separately or collaboratively, by editing agents created by others, both via structured (``CV'') an unstructured (``notes'') UIs. & Support collaborative creation of shared, custom agents for collaborative use. & Give participants the opportunity to experience and explore creating and sharing agents in a collaborative document context. \\
    \midrule
    Repeatable task list & Broadly related designs include apps for end-user development of automation rules (e.g. \textit{IFTTT}~\cite{ifttt2025}) and ``prompt libraries'' such as in \textit{Claude}~\cite{Claude2025promptlibrary}, which list prompts (text snippets) for users to copy into a chatbot conversation. & Users create tasks (via free text prompts) available for editing and use by all collaborators. They can assign tasks to agents, regardless of who had created the agent. Tasks are persisted and repeatable via automated triggers or clicking a button. & Support collaborative creation and use of tasks for agents. Support both manual and automated task triggers. & Give participants the opportunity to experience and explore creation and shared use of agent tasks in a collaborative document context, and allow them to experiment with varying degrees of agent initiative. \\
    \midrule
    Comments & Established feature in widely used products (e.g. Microsoft Word, Google Docs). & Users can write comments that tag any agent(s) to trigger them to respond. User can also ask one agent to respond to another one. Agents can also add new comments on their own in response to document tasks. & Support collaborative interaction with shared agents by integrating it into a UI element already used by teams for (human-to-human) collaboration. & Give participants the opportunity to experience and explore interaction with agents (with varying degrees of initiative) in the collaborative writing context. \\
    \bottomrule
  \end{tabularx}
\end{table*}

\subsection{Interaction with Multiple Agents}

Prior work has explored interactive systems with \textit{multiple} AI agents. \citet{Naik2025earlyadopters} interviewed early adopters to identify use cases, and \citet{Lee2025map} introduced a multi-agent system for resolving group conflicts. %
While not focused on writing, both emphasize transparency and coordination. We address these through a shared agent task list and by integrating AI output into collaborative comments.

\citet{Klieger2024chatcollab} studied collaboration in game development with three predefined agent personas. We extend this by enabling users to create their own agents through a profile UI.
Related, \citet{Siddiqui2025scriptshift} provided six fixed personas for prompting AI (``writer's friends''), while \citet{Benharrak2024aipersonas} allowed users to create many such personas. However, both projects did not consider these as agents and thus did not explore variants of initiative or integration with collaborative features.

Finally, we distinguish our work from text generation systems using multiple LLMs (e.g. \cite{Venkatraman2024collabstory}), which exclude human writers. We conceptualize ``multi-agent systems'' from the user's perspective -- as shared, interactive objects in the UI. %

\subsection{Summary}
Prior research has either focused on collaborative writing among multiple people or on writing with AI, but not on the intersection. Existing AI writing tools are overall designed for individual users and do not explore the collaborative dynamics of writing with AI, as also concluded by \citet{Chen2025nonnativecollab}. %
To the best of our knowledge, this work is the first to investigate how multiple people work together with multiple shared AI agents within a shared document environment.
\revision{\cref{tab:design_elements} summarises our key design elements in comparison to related appearances. The next section describes our design in detail.}

%% file: sections/concept.tex
\section{Concept and System Characteristics}\label{sec:concept}

We describe our overarching design goals and motivate the features implemented in our prototype. %

\subsection{Design Goals}
Our design goal is to embed AI writing support directly into a shared document environment. We realise this by introducing AI \textit{agents} into a standard collaborative editor UI. Users configure agents through \textit{profiles} and a \textit{task list}, while agent responses appear in the familiar \textit{comments} interface. To support the goal of enabling teams to use \textit{AI as a shared resource}, all profiles, tasks, and comments are visible and interactable to co-writers.

\begin{figure*}[t]
  \centering
  \includegraphics[width=\textwidth]{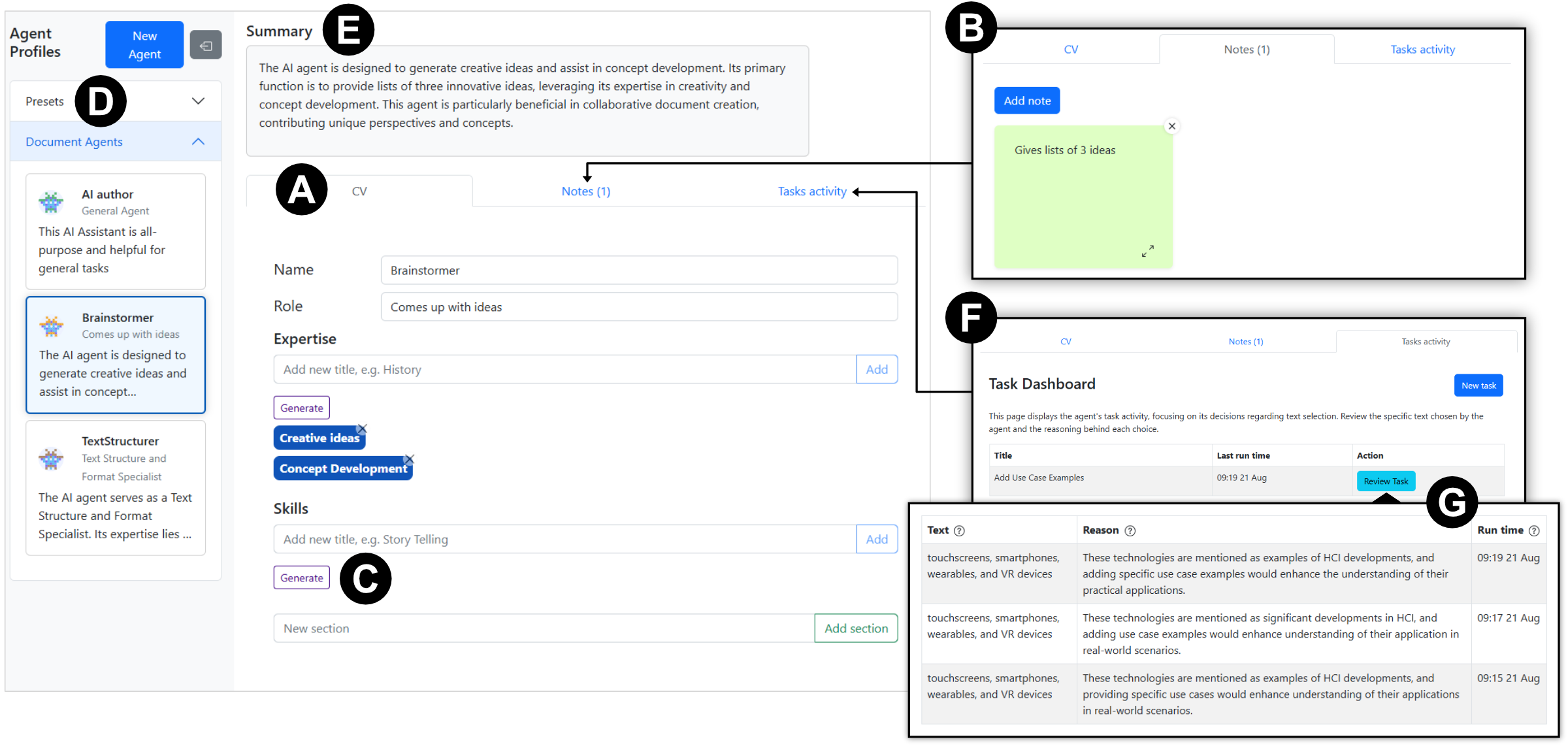}
  \caption{The UI for agent profiles and creation: Users create agents with descriptive text in two formats. In the structured format (A), they fill out a form-like ``CV'' with \textit{Name}, \textit{Role}, \textit{Expertise}, and \textit{Skills}, extendable with custom sections. In the unstructured format (B), they add free-form notes. AI suggestions for any section are available via the ``Generate'' button (C), and fields may be left empty. The UI also shows the list of created agents (D), including a list section with (editable) presets for quick setup. For each selected agent, the UI provides an automatically generated text summary (E) and a history of tasks (F). Clicking ``Review Task'' (G) reveals logs of all task runs, including the agent's selected texts, reasonings for commenting, and timestamps.}
    \Description{The figure shows the user interface for creating and editing agent profiles. On the left, a panel lists existing agents, including editable presets such as ``AI Author,'' ``Brainstormer,'' and ``TextStructurer.'' In the center, a structured form-like CV is displayed, where users can enter an agent's name, role, expertise, and skills, and add new sections. Some fields are filled with example entries like ``Brainstormer,'' ``Comes up with ideas,'' and expertise areas such as ``Creative ideas'' and ``Concept development.'' A ``Generate'' button triggers AI to suggest expertise or skills. At the top-right, an unstructured notes view shows a free-form note reading ``Gives lists of 3 ideas.'' Above the CV, the system also provides an automatically generated summary of the agent's purpose and capabilities. Below, a task dashboard displays the history of tasks completed by the agent, including a table that shows suggested text, the reasoning behind each suggestion, and timestamps for when the tasks were run.}
  \label{fig:agent_creation_ui}
\end{figure*}

\subsection{Agent Profiles}

Our design explores enabling co-writers to create multiple AI agents as part of their collaborative home~\cite{LarsenLedet2020collabwriting}.

\subsubsection{Creating multiple shared agents}
Co-writers often take on different roles~\cite{Posner1992howtogether, Lowry2004taxonomy}, motivating support for multiple agents with distinct roles. Multiple agents also let users tailor support according to different needs~\cite{LarsenLedet2021dontagree}. Our design keeps agents visible and editable to everyone to support their use as \textit{common objects}~\cite{LarsenLedet2020collabwriting}.

\subsubsection{Creating agents in structured and unstructured ways}
In our UI for agent creation (\cref{fig:agent_creation_ui}), profiles can be defined in structured form (CV-like, \cref{fig:agent_creation_ui}A), on an unstructured canvas with notes (\cref{fig:agent_creation_ui}B), or by mixing both. The structured form is motivated by work on AI personas~\cite{Benharrak2024aipersonas}, which showed that predefined fields help users get started. %
The unstructured canvas allows users to add any information.

\subsubsection{Suggestions for agent attributes}\label{sec:concept_suggestions_for_cv}
To help users define agents, the structured profile UI offers suggestions (\cref{fig:agent_creation_ui}C), motivated by findings that prompting can be challenging, in general~\cite{ZamfirescuPereira2023johnny}, in collaborative settings~\cite{Han2024teamsai}, and when describing agents~\cite{Benharrak2024aipersonas}.

\subsubsection{Agent presets}
We provide presets (\cref{fig:agent_creation_ui}D): \textit{reviewer}, \textit{idea generator}, \textit{structure \& formatting specialist}, and \textit{English teacher}. Presets help users get started, addressing challenges of prompting~\cite{Benharrak2024aipersonas, ZamfirescuPereira2023johnny} and metacognitive demands~\cite{Tankelevitch2024metacog}. %
Selecting a preset adds it to the document's agent list, where it can be edited.

\subsubsection{Generated agent summary as an overview}
Our design automatically generates a summary of each agent (\cref{fig:agent_creation_ui}E), in line with the conclusion in related work~\cite{Chen2025nonnativecollab} that AI summaries could help co-writers ``sync up'' -- in our case understanding agents created by others.

\begin{figure*}[t]
  \centering
  \includegraphics[width=\textwidth]{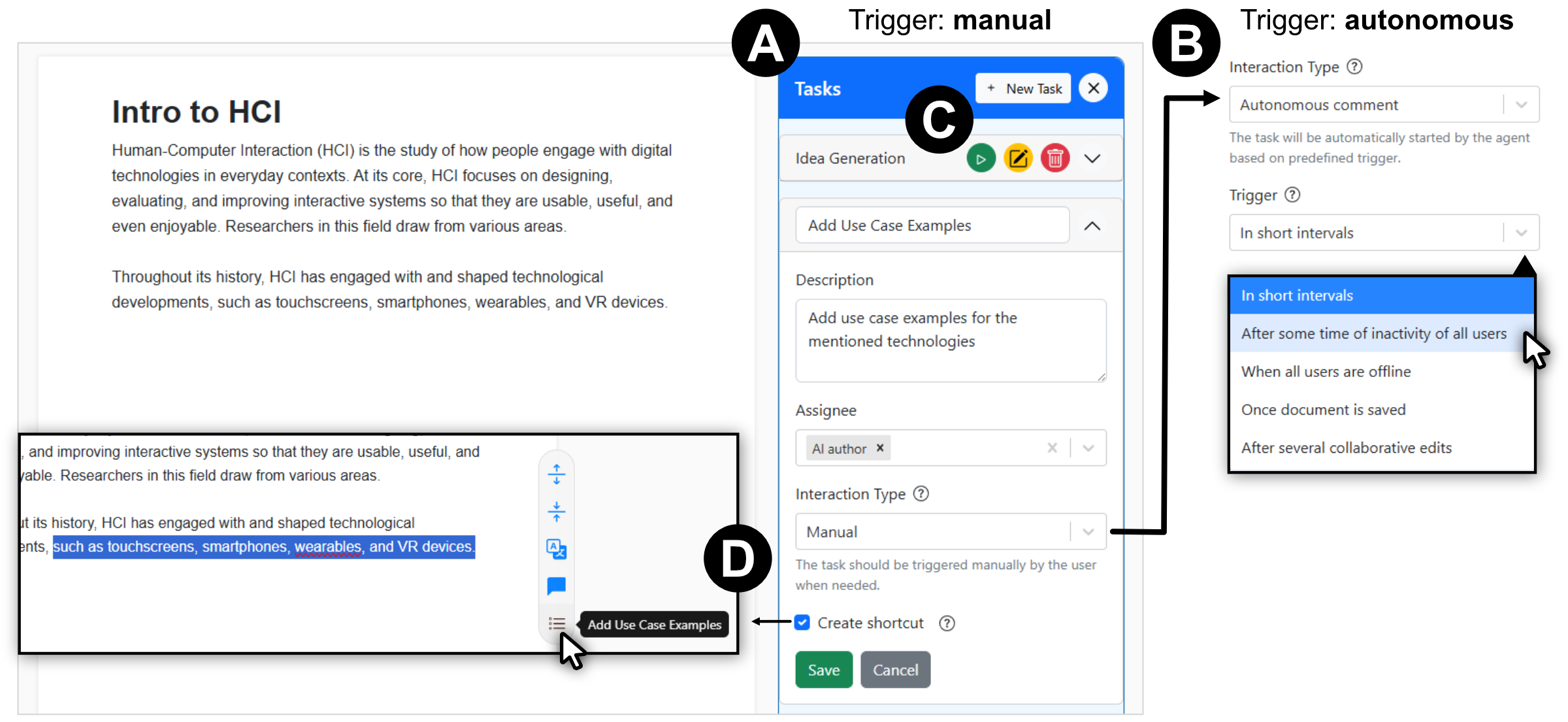}
  \caption{The UI for task creation and activation: The task list is a fold-out sidebar (A) where users create tasks for agents by entering a title, description (instruction), assigned agent (or default ``auto select''), and interaction type (manual or autonomous). For autonomous tasks (B), users choose a trigger from predefined options (elicited via a survey with academic writers, N=16). All tasks can also be run manually via the green ``play'' button (C). If ``Create shortcut'' is selected (D), the task additionally appears as a button in the floating toolbar on text selection.}
  \Description{The figure shows the user interface for creating and activating tasks for AI agents. On the left, a document is displayed with highlighted text, and a floating toolbar is shown where a shortcut button labelled ``Add Use Case Examples'' can appear. On the right, the task list sidebar is open. At the top, users can create a new task by entering a title, description, and assigning an agent, either a specific one or by leaving the default option ``auto select.'' Below, a dropdown menu labelled ``Interaction Type'' allows users to choose between manual or autonomous tasks. In the manual case, tasks must be triggered by the user, and they can also be launched using a green ``play'' button. In the autonomous case, a second dropdown menu labelled ``Trigger'' is displayed, offering options such as ``In short intervals,'' ``After some time of inactivity of all users,'' ``When all users are offline,'' ``Once document is saved,'' or ``After several collaborative edits.'' At the bottom, a checkbox labelled ``Create shortcut'' is selected, which places a task shortcut button into the text selection toolbar for faster access.}
  \label{fig:task_list_ui}
\end{figure*}

\subsection{Task List for Agents}

Task partitioning, coordination, and delegation are central to collaborative writing~\cite{Baecker1993collabwriting, Dourish1992awareness, Sharples1993collabwriting}, supported through materials such as comments and plans (e.g. todo lists)~\cite{Bodker2024material}. Users in mixed human-AI teams especially value explicit shared goals~\cite{Schelble2022thinktogether}. \citet{Tankelevitch2024metacog} identify task delegation as a metacognitive demand in working with generative AI, calling for designs to support this. These findings motivate our explicit task list for agents.

\subsubsection{Creating and editing tasks} 
Users create tasks in a fold-out sidebar next to the page (\cref{fig:task_list_ui}A), entering description (prompt), interaction type, and optionally assigning specific agents. Shortcuts in the top menu (\cref{fig:teaser}, top left) open the task list or add a new task. Existing tasks can be directly triggered, edited, or deleted via buttons (\cref{fig:task_list_ui}C).

\subsubsection{Task initiative}
Our design explores two types of task initiative: autonomous and manual. For the autonomous mode, users specify an event to trigger the task (\cref{fig:task_list_ui}B): short intervals, user inactivity, all users offline, document saved, or after several collaborative edits. We informed these options via a survey with academic writers (N=16, \revision{see \cref{sec:appendix_trigger_survey}}). In manual mode, tasks run when a user clicks the green ``play'' button in the list (\cref{fig:task_list_ui}C) or a saved shortcut (see \cref{sec:concept_task_shortcuts}). %

\subsubsection{Task shortcuts as tools}\label{sec:concept_task_shortcuts}
Tasks can be saved as shortcut buttons (\cref{fig:task_list_ui}D) in a floating toolbar, shown whenever text is selected. Thus, frequent tasks can be repeated without opening the list. Making prompts repeatable as UI elements is motivated by \textit{DirectGPT}~\cite{Masson2024directgpt}, \textit{ABScribe}~\cite{Reza2024abscribe}, \textit{DynaVis}~\cite{Vaithilingam2024dynavis}, and the toolbar idea by \citet{Dang2022howtoprompt}. As shown in \cref{fig:task_list_ui}D, our toolbar also provides four default tools (\textit{Extend}, \textit{Summarize}, \textit{Translate}, and adding a comment).

\subsubsection{Task history in agent profiles}
Whenever an agent executes a task, a log entry is added to that agent's profile (\cref{fig:agent_creation_ui}F, G). We added this for transparency \revision{(i.e. all users can see all agents and profiles)} and scrutability in linking agents and tasks \revision{(i.e. all users can inspect all task runs by all agents)}, in line with \citet{Park2023whywhy}, who found that co-writers value rationales for edits.

\begin{figure*}[t]
  \centering
  \includegraphics[width=\textwidth]{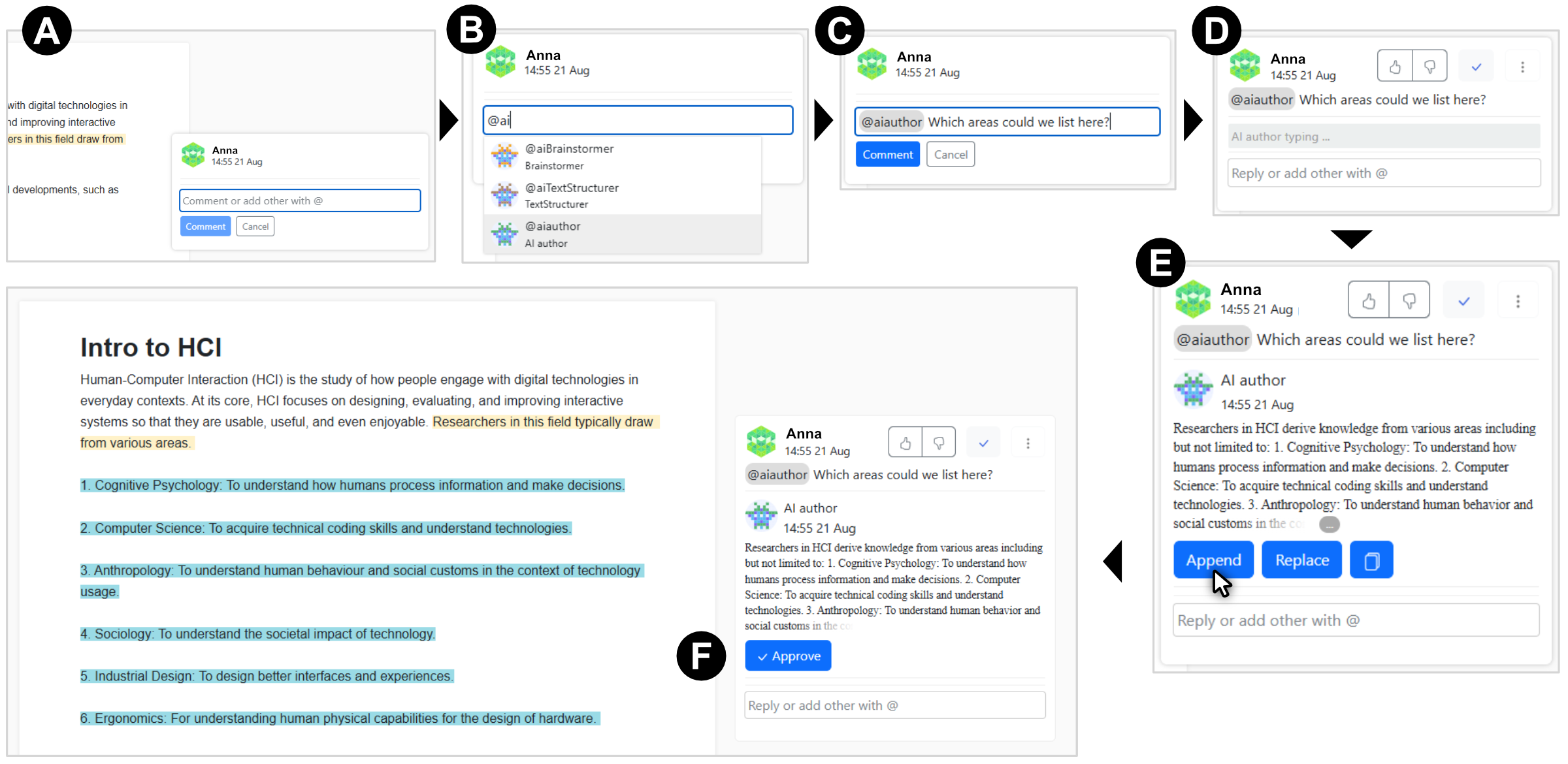}
  \caption{Example interaction flow when involving an AI agent via the comment UI: The user selects text and clicks the comment button in the floating toolbar (\cref{fig:task_list_ui}D) to open a new comment (A). By typing ``@ai'', the user brings up matching agent (or user) names (B); the user selects the default agent (``@aiauthor'') and enters a request (C). An indicator shows the agent is ``typing'' (D) while the LLM generates a response, which then appears in the thread (E). Clicking ``Append'' inserts the coloured text on the page (F), which the user edits before clicking ``Approve'' to finalize the change and close the comment. See \cref{fig:comments_ui2} for further examples.}
  \Description{The figure shows an example interaction flow where a user involves an AI agent through the comment interface. At the top, a sequence of screenshots illustrates the process step by step. First, highlighted text in the document is selected and the user opens a new comment field from the floating toolbar. In the comment box, the user begins typing ``@ai,'' which brings up a list of matching agent and user names. The default agent ``@aiauthor'' is selected, and the user writes a request, shown here as ``Which areas could we list here?'' A status message indicates that the agent is ``typing'' while generating its response. The agent's reply then appears directly in the comment thread. At the bottom, the interface shows buttons labelled ``Append'' and ``Replace,'' and the user chooses ``Append'' to insert the AI's response into the main text. The response appears highlighted in colour in the shared document. Finally, an ``Approve'' button is shown, which the user can click after making edits, finalizing the change and closing the comment.}
  \label{fig:comments_ui}
\end{figure*}

\begin{figure*}[t]
  \centering
  \includegraphics[width=\textwidth]{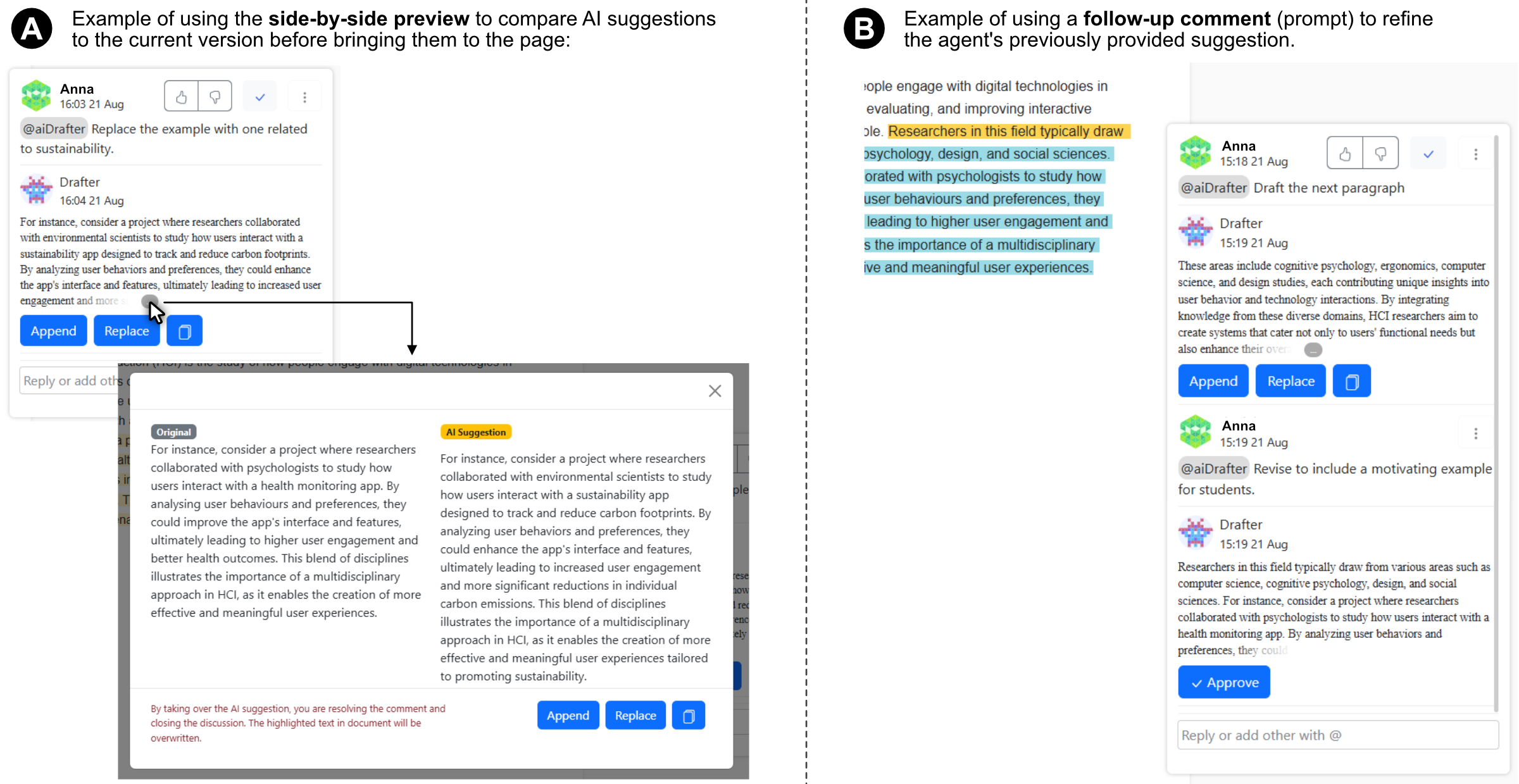}
  \caption{Two examples of further interactions with agent responses: (A) The user clicks ``...'' to view the full response in a side-by-side preview, shown next to the selected text that would be replaced if accepted. This lets users read longer AI outputs in full and compare them before committing. In (B), the user asks the agent to draft the next paragraph and follows up in the same comment thread, refining the result (here by requesting a motivating example). The screenshot shows the state after clicking ``Append'' on the refined (2nd) response, with the new text added to the page in teal colour for review and editing before final approval.}
  \Description{The figure shows two examples of extended interactions with AI agent responses. On the left (A), a user comment requests the agent to replace an example with one related to sustainability. The agent replies with a detailed text suggestion. The user clicks a three-dot button to see the full reply in a side-by-side preview window. This preview displays the original text on the left and the AI's suggestion on the right, allowing the user to compare both versions in full before deciding whether to apply the change. On the right (B), another example illustrates a multi-turn interaction. The user first asks the agent to draft the next paragraph, and the agent generates a passage. The user then follows up in the same comment thread, asking to include a motivating example for students. The agent responds with a revised suggestion. After clicking ``Append,'' the new text appears highlighted in teal within the main document, ready for the user to review, edit, and approve.}
  \label{fig:comments_ui2}
\end{figure*}

\subsection{Collaborative Comments for Users \& Agents}

Already in the 1990s, \citet{Sharples1993collabwriting} highlighted communication as a core issue in collaborative writing software, and \citet{Rimmershaw1992collabwriting} noted the value of ``meta-comments'' as annotations tied to text. Our comment design builds on these insights and established conventions (e.g. Google Docs, Microsoft Word): Users select text and click the comment button in the floating toolbar (\cref{fig:task_list_ui}D).

\subsubsection{Agents respond via comments}
Related work~\cite{Birnholtz2012tracking, Birnholtz2013writeherenow}  shows that edits carry social meaning and recommends presenting revisions as possibilities rather than definitive changes, especially when co-writers do not know each other well. This motivates our choice to have agents respond through comments, not by editing text directly. Agents select text parts to comment on via a prompt-based procedure (\cref{sec:system}). Our comment UI (\cref{fig:comments_ui}) provides buttons for users to \textit{append} a suggestion or \textit{replace} selected text, plus a side-by-side view of the original and the suggestion (\cref{fig:comments_ui2}A). Once an option is chosen, an ``approve'' button finalizes the change and closes the comment.

\subsubsection{Referencing others and AI in comments}
Users can reference co-writers in comments via ``@'' + user name, or agents via ``@'' + agent name. A generic default agent (``@aiAuthor'') provides support without first creating a bespoke one (\cref{fig:comments_ui}A-D). The referenced agent replies in the comment thread (\cref{fig:comments_ui}D, E), which co-writers can continue as usual (\cref{fig:comments_ui2}B). Further agents can also be referenced, to ``converse'' with each other.

\subsubsection{Ad-hoc tasks in comments}
Referencing agents also enables writers to delegate something directly in a comment (\cref{fig:comments_ui}), bypassing task creation and task list. This mirrors using comments for delegating tasks to co-writers~\cite{Bodker2024material}.

%% file: sections/system.tex
\section{System Implementation}\label{sec:system}
Our system consists of a backend and frontend and connects to an LLM provider.

\subsection{Frontend Application}
Our frontend uses \textit{React} with \textit{React Bootstrap} and \textit{React Quill} for the editor. We developed a custom Quill plugin for creating annotations. Real-time collaboration and syncing is implemented with \textit{Yjs}, a conflict-free replicated data type (CRDT)~\cite{yjsdocs}. We used \textit{RTK Query} for data fetching and local state management.

\subsection{Backend Application}
Our \textit{Python} server is built with \textit{FastAPI}, launching \textit{Yjs} and \textit{Socket.IO} websockets to manage real-time data updates. A second server implements a \textit{Celery} task queue for expensive background operations (i.e. agents' document tasks). %
The servers connect to a \textit{MongoDB} (e.g. persisting documents, texts, comments, agent profiles).

\subsection{AI Integration}

Both servers call an LLM API. Our prototype used OpenAI's ``gpt-4o-mini''. \revision{We chose this model to represent LLM capabilities in our study in line with a currently widely used product (ChatGPT), with fast response times.}
This is integrated into the backend via the \textit{openai} and \textit{autogen} libraries. The first is used for simple prompt queries, the second handles multi-turn conversations and agent-based interactions. The LLM is employed in the following cases:

\subsubsection{Comment response generation}\label{sec:system_ai_comment_generation}
When users mention agents in a comment, an \textit{autogen} group chat is created. The tagged agents are attached to the conversation and initialized with their profile (prompt in \cref{lst:agent_prompt}). Document text and goal, selected text, and conversation history for the selected text (if present) are added.

\subsubsection{Agent CV suggestions and summary generation}

For CV suggestions (\cref{sec:concept_suggestions_for_cv}), \textit{autogen} is queried using the prompt in \cref{lst:agent_suggestions_prompt}. %
Each time an agent is saved, its summary is updated via the \textit{autogen} agent, using the prompt in \cref{lst:agent_summary_prompt}.

\subsubsection{Document task title generation and automatic assignee selection}

When a document task is saved, a title is generated based on its description using the \textit{openai} API (prompt in \cref{lst:task_title_prompt}).
If the user does not specify an assignee, the system queries \textit{openai} (prompt in \cref{lst:task_assignee_prompt}) to select one based on task and agent descriptions. The LLM returns a confidence score; if it is 85\% or higher, the recommended agent is assigned. Otherwise, the system assigns the default agent.

\subsubsection{Task execution across the document}

When a task is executed on the entire document, the \textit{Yjs} websocket server sends the task to the \textit{Celery} server. Its response is then applied to the \textit{Yjs} document and stored in the database. In detail:

\begin{enumerate}
    \item \textit{Generating selection proposals (Celery):} The first agent assignee selects relevant text segments. Based on the assignee and task description, the LLM is queried to return the segments, their length, reasoning, and confidence score (prompt in \cref{lst:task_text_selection_prompt}).
    \item \textit{Filtering selection proposals (Celery):} Segments overlapping with existing text annotations or scoring below 80\% confidence rate are discarded. The system notes attempted execution in the existing text annotation comment.
    \item \textit{Generating responses (Celery):} For each valid segment, a conversation with the task assignees is initiated using the process as in \cref{sec:system_ai_comment_generation}. Results are returned to the \textit{Yjs} websocket server.
    \item \textit{Integrating responses (Yjs websocket):} New annotations and comments are integrated into the \textit{Yjs} document, while existing annotations are updated. %
\end{enumerate}

\subsection{Autonomous Tasks}\label{sec:system_autonomous_tasks}

The \textit{Yjs} websocket server continuously monitors document activity and evaluates the trigger conditions (\cref{table:time_triggers_implementation}), to request task execution from the \textit{Celery} server. %

\begin{table}[t]
  \small
  \caption{Autonomous task triggers and their implementation.}
  \label{table:time_triggers_implementation}
  \Description{A two-column table listing triggers for autonomous tasks and the corresponding conditions that cause each trigger to fire.}
  \centering
  \renewcommand{\arraystretch}{1.6}
  \setlength{\tabcolsep}{4pt}
  \begin{tabularx}{\columnwidth}{@{}>{\raggedright\arraybackslash}p{0.3\columnwidth} >{\raggedright\arraybackslash}X@{}}
    \toprule
    \textbf{Trigger condition} & \textbf{Implementation} \\
    \midrule
    In short intervals &
    As soon as someone enters the document socket room, schedule a job to run every 5 minutes. Skip if a job is already scheduled. \\
    After some time of inactivity of all users &
    Debounce 2 minutes after the last detected user activity (text or comment updates). \\
    When all users are offline &
    When the last online user disconnects from the document socket room. \\
    Once document is saved &
    When a user clicks the \textit{Save} button. \\
    After several collaborative edits &
    The \textit{Yjs} document includes a shared array that tracks the users who typed in the document. If the list has more than two users, it indicates a collaborative edit. The array is cleared after the trigger is executed to detect future edits. \\
    \bottomrule
  \end{tabularx}
\end{table}

%% file: sections/user_study.tex
\section{User Study}

We conducted an exploratory user study with our prototype as a \revision{technology} probe to investigate how co-writers collaboratively create and interact with AI agents. 
\revision{We motivate our approach by following the mapping of the three goals of technology probes~\cite{Hutchinson2003techprobe} to LLM-involved design research by \citet{Feng2025canvil}. 
A main interest with our study maps to the \textit{social science goal}: understanding how teams experience the integration of (customisable) AI agents into their shared documents. %
Another interest maps to the \textit{design goal}: ``inspiring users and researchers to think about new technologies''~\cite{Hutchinson2003techprobe}, in our case collaborative integration of agents in shared documents. 
In addition, our probe serves the \textit{engineering goal} of field-testing, for a prototype release to facilitate future studies.}

\subsection{Participants}

We recruited \studyN{} participants (14 women, 16 men) in \groupsN{} groups by sharing a study call in our network across a few universities, where we reached mostly faculty and students (\cref{table:method_participants} in Appendix). %
Ten groups signed up directly, three signed up individually but knew each other, and one group (G7) was formed by us from two individual registrations.
Participants' age ranged from 22 to 56 years (median 27), and they all had sufficient knowledge of English. 
Their median frequencies of using LLMs and writing documents was daily, and collaborative writing was weekly. Their writing domains included academic study, professional work, research, and journalism.
They were compensated with \SI{25}{\text{€}}.

\subsection{Study Procedure}

Once participants received the study link and provided informed consent, they were free to work with the prototype for a week. We recorded interactions and conducted semi-structured group interviews using Microsoft Teams, which were audio-recorded and transcribed. The study followed our institute's regulations (e.g. information on goals, process, data protection, opt-out and consent form). %

\subsubsection{Introductory meeting (15 minutes)}
Based on feedback from the first three groups, we concluded that a quick tour of the features would be helpful. %
Thus, we conducted online introductions for the remaining groups, where one researcher demonstrated the features via screen sharing.

\subsubsection{Writing with the prototype (7 days)}
The procedure was coordinated via the \textit{StudyAlign} web application \cite{Lehmann2025_studyalign}, which integrated the questionnaires, writing task and description page, and access to the prototype. 
\begin{enumerate}
    \item \textbf{Introduction questionnaire (5 minutes):} It covered demographic information and writing experience. %
    \item \textbf{Task briefing (20 minutes):} Participants chose between two tasks -- either working on a real task of their own, or, if none was available, writing an essay on positive and negative aspects of AI in daily life. The briefing explained the prototype's features through text and videos.
    \item \textbf{Writing task (7 days):} Participants used a join code to access their shared document. Groups were free to decide how to approach the task and collaboration.
    \item \textbf{Final questionnaire (25 minutes):} Participants answered SUS~\cite{sus_eval} and CSI~\cite{csi} questionnaires, evaluated individual features, and provided comments on pros and cons.
\end{enumerate}

\subsubsection{Semi-structured interview (30 minutes)}
Group interviews via video calls included questions on likes/dislikes, collaboration, challenges, agents, tasks, and triggers. This group format allowed teams to reflect on their experience together. %

%% file: sections/results.tex
\section{Results}

\revisionCR{We first report results from logging data and questionnaires (\cref{sec:results_interaction}) as a summary of interaction in the study, before turning to our main qualitative analysis (\cref{sec:results_qualitative}).}

\subsection{Usability and Interaction Behaviour}\label{sec:results_interaction}

\revision{While our focus is on qualitative exploration, we also} logged interactions \revision{to support and contextualise the qualitative data}. One group (G4) accidentally accessed the prototype outside of our study setup, such that logs were not recorded. We reconstructed their interactions from our database where possible and note where they are excluded.

\subsubsection{Usability ratings}\label{sec:results_usability}

\revisionCR{To provide basic background information,} participants initially rated how often they work on texts and how they collaborate (\cref{table:method_participants} in the Appendix). %
Further tools used when collaborating on documents included Slack, Mattermost, and Teams (\cref{fig:ratings_frequency_collaboration} in the Appendix). Sixteen use such chat tools daily, six weekly, four monthly, and four less than monthly.

\revisionCR{We also assessed usability measures as an overview of participants' perception of the design features, which we enrich per feature in our qualitative analysis in \cref{sec:results_design_feature_feedback}.} Ratings on the system usability scale (SUS) had a mean score of 66.92 (SD=12.96, min=45, max=90), see \cref{fig:ratings_sus} in the Appendix. This score is slightly below the average benchmark of 68. %
The CSI yielded a mean of 67.21 (SD=15.60, min=31, max=91). As shown in \cref{table:csi_factors} in the Appendix, the most valued factors of the CSI were ``Enjoyment'' and ``Collaboration''. 
We further asked Likert items on general feedback (\cref{fig:ratings_general_feedback}) and agents (\cref{fig:ratings_ai_agents}). Most found the interaction natural and felt it helped them \revision{reach their goals faster}. Around a quarter (23.33\%) did not feel in control of the text. The agent features were rated positively overall but participants were divided about whether one agent is enough or not.

\subsubsection{Metrics on agent creation}\label{sec:results_metrics_agent_creation}

\revision{Participants created} 39 agents, 2.79 per group (SD=1.25, min=0, max=4). %
38.46\% were created in the first quartile of each group's normalised editing time, 28.2\% in the second and third, and 33.33\% in the last. \revision{This reveals: (1) initial exploration, typical for experiencing new features -- yet not exclusively, suggesting (2) emerging needs or preferences addressed with agents created later. A variety of needs is also reflected in the text analyses of profiles and tasks in \cref{sec:results_further_text_analyses}.} %

The following results exclude G4.
Agents could be defined through the ``CV'' and notes features (\cref{fig:agent_creation_ui}): 54 CV sections were created (28 skill sections and 26 expertise sections) with 71 values added manually (33 skills, 38 expertises). Responding to participants' requests for suggestions, the system suggested 174 values, and participants added 45.4\% (79) of those (36 skills, 43 expertises). Notes were added to custom agents in four cases only. Another three notes were added to agents extended from presets. \revision{These results show that} people mainly used the structured UI (``CV'') for defining agents. \revision{In contrast, the free-form notes were not seen as adding further value.}

Once created, agents were updated in eleven cases; ten times by the agent's respective creator, and only once by another team member. \revision{Together with the qualitative results (\cref{sec:results_territory}), this shows} that agents were seen as belonging to their creator.

\subsubsection{Metrics on tasks from the task list}

Participants created 67 tasks via the task list feature (\cref{fig:task_list_ui}), 5.15 per group (SD=2.19, min=1, max=9), excluding G4. Most (88.06\%) were created by the author of the agent to which the task was assigned.
The tasks from the task list were triggered manually 113 times; %
84.7\% by the task's creator.
Task editing occurred 10 times. 
\revision{For tasks with autonomous triggers (7 tasks), the most common trigger was ``After some time of inactivity of all users'' (3 tasks), followed by ``In short intervals'' (2), and ``Once document is saved'' and ``When all users are offline'' (1 each).}

We also analysed task creation across the normalised editing time per group. Most tasks were created in the last quartile (41.67\%) and in the third (23.33\%), \revision{compared to} the first (21.66\%) and second (13.33\%). \revision{Combined with the agent results, this indicates that custom agents are created before custom tasks. Participants' rationales for valuing custom agents explain this: Some saw agent profiles as a way to ``preload'' context for later task prompts (\cref{sec:results_one_vs_multiple_agents}).}

\subsubsection{Metrics on comments}\label{sec:results_metrics_comments}

We counted 468 comments; 376 were added by agents in response to running tasks, while participants added 92 comments. G2, G5 and G12 never used comments for ad-hoc task delegation to AI. The other groups created a mean of 8.36 comments (SD=7.19, min=1, max=22). \cref{table:discussion_patterns} shows specific discussion patterns, revealing that only short discussions happened in comments. %
\revision{For contextualisation, note} that most groups reported to use further communication tools, explaining the small number of human-human discussions \revision{via comments here}.

\begin{table*}[]
\newcolumntype{L}{>{\raggedright\arraybackslash}X}
\caption{Identified discussion patterns from analysing comment and reply events.}
  \label{table:discussion_patterns}
  \Description{A four-column table listing the five discussion patterns that appeared in the log data. Each pattern is described and has a count.}
\begin{tabularx}{\linewidth}{lLLr}
\toprule
\textbf{Discussion pattern} & \textbf{Description} & \revision{\textbf{How this was used}} & \multicolumn{1}{l}{\textbf{Count}} \\ \hline
human-human & comment by user, reply by team members & \revision{coordinating tasks} & 4 \\
human-AI & comment by user, AI replies; single-turn & \revision{delegate task; directly accept/reject AI suggestion} & 57 \\
human-AI negotiation & comment by user, AI replies, human replies; multi-turn & \revision{refinement of AI suggestions} & 17 \\
human-team-AI & comment by user, AI replies, team members reply; multi-turn & \revision{work together on goals (e.g. brainstorming); involve different agents} & 6 \\
AI-human & comment by AI, at least one user replies & \revision{refinement of AI initiated suggestions} & \revision{6} \\
\bottomrule
\end{tabularx}
\end{table*}

To accept agents' suggestions in comments, we offered three methods: %
Appending text at the end of the selected text, replacing the selected text, and copying the suggestion to the clipboard. %
Excluding G4 again, suggestions were \textit{appended} 76 times; 50 times by the comment's creator, 26 times by others (19 times when the comment was created by a user, 7 times when it was created by AI). \textit{Replace} was used 60 times; 29 times by the comment's creator, 31 times by others (23 times for user-created comments, 8 times for AI). \textit{Copy to clipboard} was only used 16 times: 9 times by the comment's creator, 7 times by others (6 times for user-created comments, once for AI). Thus, agent-initiated suggestions were less often accepted than user-initiated ones.
\revision{This had two reasons: (1)~User-initiated AI suggestions reply to a user's comment on selected text and thus respond at the user's current focus, making acceptance more likely (``as is'' or for further editing them). (2) In contrast, agent-initiated comments from running a document task can result in a high number of comments (anywhere); when feeling overwhelmed, users ignored some of them. See \cref{sec:results_design_feature_feedback} for details on this from the qualitative insights, and \cref{sec:results_metrics_collab} for further related logging analyses.}

We analysed when participants posted comments: 4.83\% of comments were posted in the first quartile, 21.38\% in the second, 11.03\% in the third, and 62.76\% in the last. This indicates that participants \revision{typically} sought support from agents after writing a first draft. %

\subsubsection{\revision{Metrics on collaboration}}\label{sec:results_metrics_collab}

\revisionCR{Here we provide two metrics on collaboration as contextualisations for the qualitative results on how teams collaborated in our study.} \revision{These analyses exclude G4.}

\paragraph{\revision{Interaction timing and engagement with (each others') AI comments}}
\revision{We computed the times from AI adding a comment until a user accepting it. The grand mean was 19.98 minutes (SD=121.41, min=0.01, max=1363.53). Users accepted AI suggestions responding to human comments (136) after 22.26 minutes (SD=128.21, min=0.07, max=1363.53). They accepted AI suggestions from AI-initiated comments less often (16) but faster, after 0.67 minutes (SD=0.70, min=0.01, max=2.07). The difference is partly due to acceptances by group members with a long time gap, such as when working asynchronously. Concretely, users accepted 48 suggestions made by AI in response to another user's comment, with a mean of 59.56 minutes (SD=212.06, min=0.14, max=1363.53). They accepted (88) AI suggestions in response to their own comments faster, with a mean of only 1.92 minutes (SD=5.08, min=0.07, max=34.61). Thus, AI comments became more temporary suggestion displays for a single user when integrated into the requesting user's own workflow, whereas they remained (longer lasting) action items when left for others (cf.~\cref{sec:results_collab_processes}).} %

\paragraph{\revision{Synchronicity with each other and AI}}
\revision{We measured interaction switches to analyse how interleaved each group collaborated. We ordered input events chronologically and counted switches of author IDs per group. %
We also included agents via AI comment events and acceptance events. The grand mean was 219.07 interaction switches (SD=285.28, min=10, max=990). See \cref{table:interaction switches} in the appendix for details -- the numbers, both absolute and relative (switches to human vs AI), help further characterise our sample: We observed a range of collaboration styles, with some groups collaborating with highly intertwined edits (e.g. G1, G14), and others more asynchronously or in sequence (e.g. G8, G11). The same holds for working with the agents, with which some groups had more back-and-forth ``handovers'' than others. \revisionCR{We see these results as another way of characterising the diversity of workflows in our study and focus on the qualitative data for further insights into these collaboration patterns (\cref{sec:results_collab_processes})}.} %

\subsubsection{Final texts}

The final texts had a mean length of 1623 words (SD=1306.93, min=226, max=4767). To assess the readability of texts, we computed the Flesch reading-ease score~\cite{Flesch1948readingease}. We removed four texts for this analysis since they included non-English parts. %
The mean readability score of 17.71 (SD=13.71, min=2.87, max=38.65) can be interpreted as ``very difficult to read''. \revision{For contextualisation, these scores fit to those reported in analyses of contemporary academic literature~\cite{PlavenSigray2017readability, Yeung2018readability} and are described as appearing in ``Scientific'' and ''Academic'' texts~\cite{Flesch1948readingease},} reflecting the nature of most of the groups' writing projects.

\subsection{Interviews and open comments}\label{sec:results_qualitative}
We transcribed the interviews with Microsoft Teams and analysed them following open and axial coding principles~\cite{Corbin2014basics}, verifying transcripts against audio. One author iteratively created, grouped, merged, and split codes into (sub)categories. Another reviewed and refined this, going back to the recordings where needed and involving the team for consensus. We similarly coded open comments from the final questionnaire. %

\subsubsection{Territorial functioning: An agent belongs to its creator}\label{sec:results_territory}
Agents and tasks were seen as belonging to their respective creators, with all groups except for two expressing this. %
One rationale \revision{for this} is that \emphresult{everyone knows best what they want from AI support}; as \p{2}{4} said \pquote{we had two or three different agents, so I had one more which I wanted to focus more on [...] fixing typos or grammar.}
And \p{1}{5} found that \pquote{you cannot really decide how many AI assistants someone would need}. Similarly, \p{2}{2} explained: \pquote{We have different AI agents and we can set different tools and [...] we can use them for a different purpose.}
\p{1}{7} valued this for prompting: \pquote{I know what to expect. It's better to use an agent that I made myself, that I know [...] So I also then know how, or I have a better feeling over how, I write the prompts in the comments.} 
\p{2}{13} even envisioned individual visibility: \pquote{Something that might be useful for one author might actually be a hindrance to another author, so [...] people could have their own agents that only show up in their version of the document or on their screen.}

Another rationale is \emphresult{controlled customisation and representing a specific user}. As \p{1}{2} said: \pquote{AI may not always match individual writing styles or preferences.} Related, \p{2}{7} wanted to ensure own agents stay as defined: \pquote{I would not like [others to edit it] directly if I'm not seeing this, and then I [...] come to my agent and he's not, like, the same.} \p{1}{8}, \p{2}{10}, \p{1}{11}, and \p{1}{12} expressed similar sentiments and \p{1}{10} framed it in terms of identity: \pquote{I considered his agents to be him. Kinda. [...] When I used the AI, I was still, like, this is still me.}
Even group 14, who stated to see agents as shared, envisioned explicit ownership controls: \pquote{Something to consider is that you have some form of rights or access by the owner} (\p{2}{14}) and their partner \p{1}{14} added that \pquote{each of us have our own parts in the text and we want to keep our style.} Group 12 also envisioned such controls.

Finally, \emphresult{workflows} \revision{that teams approached the study with} were another reason.
As \p{2}{11} said: \pquote{I have certain ways of organising myself [...] so therefore I think it would be [better] for my workflow.} Group 12 expressed a similar view %
and group 1 described they explored the new features individually. %

In summary, we found that people stayed away from iterating on the profiles of agents not created by themselves.

\subsubsection{Emerging roles and norms: Sharing agents is caring}\label{sec:results_sharing_agents}

The sense of ownership extended only to the agent profiles and the \textit{creation} of agents/tasks. In contrast, \emphresult{it was accepted to \textit{use} others' agents} (e.g. ask them in comments). With three exceptions, all groups explicitly commented on this in the interviews. 
As \p{1}{9} reflected: \pquote{I did feel comfortable using all people's agents, but I did not want to update [them].}
And \p{1}{6} said: \pquote{We tried to use all the agents [...], even if it was created by [another person].}
Similarly, \p{2}{4} remembered: \pquote{The language agent was created by the others and I used that as well. [...] I created two own tasks and just used the ones from the others as well.} 
Related, \p{2}{3} remembered using comments of others' agents: \pquote{When I opened the document, there were [...] suggestions [...]. So I read them and I accepted some of them and integrated [them], and some of them also, I think, I deleted them.}

People \emphresult{created agents and tasks with collaboration in mind}. 
For example, \p{1}{8} was keen to \pquote{make sure that the agent in the end is reliable and it's a good team player for both of us.} %
And \p{2}{2} described it for tasks: \pquote{We didn't split it one task per person. We just kind of mix it.} 
\p{2}{11} saw individually owned agents as an optional team offer: \pquote{I'd be happy to use shared ones, but [...] if I don't like them, then I still want to have my own. [...] I'd be happy to share my stuff and people should only use it if they find a benefit.}
And \p{1}{7} reflected: \pquote{As a team, you could have this agent. Everybody knows what it's for and you know you use it for this task. So of course, not everyone has to make their own five agents.}

In summary, once agents and tasks were created, teams thought of them as a shared resource for collaborative use.

\subsubsection{One vs multiple agents: A tradeoff between expected functional and perspective values}\label{sec:results_one_vs_multiple_agents}
We observed a specific strategic tradeoff: People weighed the setup effort against the expected benefits of multiple agents. Different views were discussed within nine of the teams. Overall, 14 participants brought up rationales in favour of one agent, and 20 (also) in favour of multiple ones. \revision{We report on their rationales in detail next:}

\paragraph{\revision{Rationales for one agent}}
\emphresult{Those who preferred a single agent focused on the AI's functional value}. Many considered the functional \textit{scope}, such as \p{2}{5} (\pquote{One AI agent is enough, it could make all tasks.}) and \p{2}{7} (\pquote{[The default agent] also understands and will do the task}). \p{2}{1} reflected on \textit{efficiency}: \pquote{I don't want to spend my time [...] creating another agent and giving him the background, when I know that the first agent already exists.} %
And \p{2}{3} considered \textit{creative effort}: \pquote{I had no idea of which agent I need.} %

Another rationale was that \emphresult{differences between agents were not clear enough}. As \p{1}{4} said: \pquote{As the agents [...] were quite similar in what they were doing, so, one agent would be sufficient for us.} Similarly, \p{1}{6} found that agents \pquote{did not differ from each other that much. [...] The core skills are the same.} \p{2}{7}, \p{1}{10}, and group 11 made similar statements.

\paragraph{\revision{Rationales for multiple agents}}
\emphresult{Those inclined to use multiple agents focused on perspective value}, such as \p{1}{12}: \pquote{I also really liked using AI agents to incorporate viewpoints that are not my own.} Similarly, \p{1}{1} said: \pquote{You can use the second agent to check the text, like, to play as a professor, as a reviewer.}
Group 12 also liked agents as personas for feedback. 
\p{2}{4} liked \pquote{having different agents for different things}, echoed by \p{2}{13} and \p{2}{14}.
\p{2}{8} compared it to human teams: \pquote{When you are collaborating with others in the same team, you know, some people are more good at this. Some people are more good at that. So for agents it's the same.} %
\p{2}{10} reflected on both value and effort: \pquote{I can have an external reviewer that reviews from the lens of that community. [...]  I'm a bit sceptical [about] having to create and manage and stuff.}

A related rationale was that \emphresult{agent roles support prompting}: \p{2}{7} said that \pquote{it could be faster then to refer to specific agents [...] and not explain the only one AI agent [everything]. It's like working in a company. Everybody has its role.} Similarly, \p{1}{3} considered that defined roles shorten task prompts: \pquote{It can be beneficial to have several agents predefined because their answers can be different. Sometimes I want to create a very, very specific description. I don't want to write a very long task for that.} \p{1}{7} also found that \pquote{defining the agent spares you a part of the prompts writing.} And \p{1}{13} said: %
\pquote{I first thought of the persona pattern [for prompts] and I would try to push the LLMs as much as possible into this role. [...] I thought of tasks I would assign to the agents and design the agents the way I want to implement the tasks.}

\paragraph{\revision{Summary}}
Participants weighed setup effort against benefits of multiple agents. \revision{What drives their varying preferences is their individual focus on either functional or} perspective values and future prompting. \revision{Notably, people did \textit{not} mention specific writing use cases in their rationales (e.g. one agent for emails, while a paper needs more), suggesting that they made these considerations on a more general level of what they expect from AI in writing.}

\subsubsection{\revision{Collaborative processes with AI agents}}\label{sec:results_collab_processes}

\revision{Eight groups worked synchronously, eight (also) asynchronously. The groups also used chat apps and video calls. Six groups split the writing by ``location'' (e.g. one section per person). Thus, our sample covers using agents, tasks, and comments in various collaborative processes.}

\paragraph{\revision{AI comments as team ``action items''}}
\revision{Consistently throughout the groups, AI comments became shared material (\cref{sec:results_sharing_agents}), as participants applied or rejected AI suggestions in comments, regardless of who had created or triggered the agent (\cref{sec:results_metrics_collab}). Thus, these \textbf{decisions on AI suggestions became a new, shared ``action item''} in the collaborative process. This was an overarching aspect, not tied to a specific moment or stage of writing.}

\paragraph{\revision{Using tasks for planning and structuring}}
\revision{The shared task list (\cref{sec:results_sharing_agents}) was integrated into collaborative processes in a forward-thinking way: For example, G1 created tasks for planning, while working synchronously. In contrast, G11 worked asynchronously and in the interview reflected on using tasks for structuring a document first (\revisionCR{\pquote{For this overall structure [...], top down breaking things down [...] I think tasks would work better.}, \p{2}{11}}). And for G12, who worked asynchronously and also in parallel, the tasks for agents felt similar to writing todos to each other. Thus, the \textbf{task list feature was integrated into team processes with a forward-thinking perspective}, in contrast to the more ad-hoc comments.}

\paragraph{\revision{Agents as ''glue'' for collaborative processes}}
\revision{A few people supported asynchronous collaboration with agents and tasks. 
For example, \p{2}{4} used the all-users-offline task trigger to see recaps via AI comments when coming back: \revisionCR{\pquote{After logging in again it created the summary which was quite nice. So just to get a recap of what had happened so far.}}
\p{2}{12}, also working asynchronously, reflected on interaction with agents as available alternatives for delegating tasks: \revisionCR{\pquote{I find it really useful [...] because I think it's a natural way when you work together [...] Yeah, you write a task for another co-author [...] and so I think I made no difference if I assign a task to [him] or to an AI agent. I think it was, from the feeling, nearly the same.}}
\p{1}{7} used an agent to keep work coherent that they had split across sections with their partner: \revisionCR{\pquote{I've read what he's done and then I could come back and then ask [the] agent: He wrote something like this. Can you put something related now in my text?}} %
And in G5, who split the work into sections to work on asynchronously, \p{1}{5} created a ``praiser'' agent for motivation in the absence of their partner. 
In summary, in such cases, the agents were used to \textbf{support ``handover'' or ``partner absent'' moments of asynchronous workflows}.}

\paragraph{\revision{Using (custom) agents in roles specific to writing stages}}
\revision{Many considered agents to address specific stages in their collaborative workflow. For example, G2 used them to create a first draft synchronously, while G3 and G4 asynchronously created custom ``reviewer'' agents, as also reflected on in G10 \revisionCR{(\pquote{I can have an external reviewer that reviews from the lens of that community}, \p{2}{10})}. G8 designated one user to setup all agent roles for the team. In such cases, (custom) \textbf{agents were brought in with awareness of the team's (negotiated) workflows and with writing process stages in mind.}}

\paragraph{\revision{Concerns about workflows with agents}}
\revision{Some reflected on downsides for collaboration: G10 commented that agents cannot take responsibility for mistakes. \p{2}{8} felt they interacted more with agents than their partner at times, despite sharing a video call \revisionCR{(\pquote{It feels like I'm still more, like, interacting with the AI agents, rather than the people})}. \p{2}{13} expected conflicts when one user deactivates an auto-triggered task that another finds helpful. And G4 used the ``all users offline'' trigger to avoid getting interrupted by agent comments with other auto-triggers. In summary, the concerns reveal reflections on possible \textbf{distractions} and \textbf{interruptions}, as well as questions of \textbf{responsibility}.}

\subsubsection{\revision{Values and demands of the design features}}\label{sec:results_design_feature_feedback}

\revision{Participants' rich feedback, combined with logging data and ratings (\cref{sec:results_usability}, \ref{sec:results_interaction}), enables us to dissect the values and demands of the features.} %

\paragraph{\revision{Value: Integration compared to known designs}}

Participants %
valued the direct editor integration. %
\revision{Their comparisons to their usual work highlight two aspects:}
\revision{One valued aspect is \emphresult{avoiding the need for switching between the writing environment and AI}, including copy-paste:
\p{1}{5} found the \pquote{assistants were pretty useful [...] Previously, I had to divide my text into several fragments if I wanted to check it in ChatGPT.} 
And \p{3}{4} found it \pquote{more comfortable than copying the stuff over to, for example, ChatGPT.}
\p{1}{9} also shared this value: \pquote{You can use AI agents to brainstorm new ideas. You don't have to go to GPT.}
}
\revision{A related aspect is that this \emphresult{integration is perceived as fast}, in comparison to separate AI tools:
As \p{2}{3} said: \pquote{I could do everything much faster without using ChatGPT or other applications [...] because I could use these agents.}
And \p{1}{2} had \pquote{a great experience to use it directly in the document, and it saves a lot of time [and] it was surprisingly good.} 
\p{2}{7} shared a similar view: \pquote{What I like about it, it was helping to make it really faster than I will do it without it.}
And \p{2}{1} said \pquote{it saves a lot of time that I don't have to switch tabs. I often use [...] other online tools. And in this app [...] I could just ask the agent.}
\p{2}{12} reflected on both aspects: \pquote{Previously my workflow was that I take my text, copy it to ChatGPT or other chatbots, and write, each time again, my prompt or copy the prompt, and ask GPT to do that. And with the agent it's much easier and faster because I can [...] work with different agents in parallel. That I found really, really helpful.}
}

\paragraph{\revision{Value: Using comments with agents}}

\revision{The design \emphresult{integrating agents specifically into comments was valued}, as explicitly highlighted by} \p{1}{1}, \p{1}{3}, \p{2}{9} and everyone in groups 4, 8, 11, 12, and 13. \revision{For example,} \p{2}{8} said: \pquote{I really like the comments. You can add the agent directly in the comments and that was very convenient, and it feels like you're actually talking to a person or like you're discussing with someone.} Similary, \p{2}{3} found the comments \pquote{very helpful, especially to collaborate with others [...] and very integrated.} 
\revision{\p{2}{11} valued the AI comments as additions in comparison to Google Docs: \pquote{I like the overall [...] similarity to Google Docs style because, I mean, for me it was very easy to to get into it [...] and then having AI integrated in the form of [...] comments and to have these tasks. [...] I really like to have AI integrated in the tool.}}
\p{1}{7} wanted to keep this feature \revision{when reflecting on their usual tools}: \pquote{What I would include in the usual editor is the comments with the AI agents.}

\paragraph{\revision{Demand: Setup of agent profiles}}

For some, \emphresult{using the profile UI was easier than figuring out \textit{what} information to enter.} As \p{1}{6} said, \pquote{AI agent creation [is] easy to understand} and \p{1}{11} found it \pquote{really easy to get your head around, to imagine that you have this little team}, while \p{2}{4} \pquote{found it a bit confusing to decide whether it now was a skill or just a background.} \p{1}{3} reflected explicitly on both sides: \pquote{I did not struggle to create agents. Basically, because the interface for agent creation was really easy to understand, but like, some intuition about what kind of agent I want to create. That's actually kind of [the] problem.} \p{1}{13} generalised it: \pquote{It's the typical issue with LLMs. If you're not really sure what you want to generate, then the answers are typically not so high quality.} \revision{In summary, agent setup was a metacognitive demand (cf.~\cite{Tankelevitch2024metacog}) rather than a usability demand, fitting to prompting challenges in related work~\cite{Benharrak2024aipersonas, ZamfirescuPereira2023johnny}.}

\paragraph{\revision{Demand: Verbose AI output}}
\revision{The} main point of critique concerned LLM output. G10 was the most critical; it did not meet the bar for them, as they perceived it as worse than ChatGPT. All other groups found it useful overall but voiced the concern that \emphresult{AI output was too verbose}, in three flavours: 
First, AI agents added \emphresult{too many comments}, when not tagged in a specific place via a comment but rather given a task via the task list. As \p{1}{3} concluded \pquote{it's left, like, comments over all [the] document and I would not want to use these autonomous tasks too often.} 
Second, some AI \emphresult{comments were longer than desired}, which made \p{1}{4}, for example, suggest to \pquote{shorten the comments quite a bit}.
Third, AI responses included \emphresult{unwanted conversational parts} (e.g. ``Sure, here's a draft: ...''), such as in the case of \p{1}{6}: \pquote{[agents] reply in conversational messages, [which would be] added to the text.}

\paragraph{\revision{Addressing the demands and control needs with manual triggers}}

Overall, \emphresult{participants preferred to give tasks via manual triggers and comments.} 
\revision{One} reason was that \emphresult{calling agents manually in comments constrains their response} to one new comment in the expected location. \revision{This also addresses AI verbosity reported above via targeted requests.} For example, \p{1}{13}, \p{2}{4}, and \p{1}{3} highlighted how it was useful to \pquote{tag an agent and ask him a question regarding the selected text} (\p{1}{3}). %
Related, \p{2}{13} liked this for task shortcuts: \pquote{I really like the feature where you [...] select the text [...] and just use a task you defined beforehand.} 
\revision{Another} reason was the lack of progress feedback or the inherent invisibility of automated triggers such as leaving the document. For instance, \p{2}{5} said: \pquote{It works, but I have seen no results, but I have not understood why, but only yesterday, I see it was there.} %

Some reflected on \emphresult{trust and control} for preferring manual triggers, such as \p{1}{14} (\pquote{You don't have much control of what's going on.}) and \p{1}{7}: \pquote{You're always afraid that some changes will happen to your documents. [...] So I don't think people will be doing a lot of really autonomous stuff.} 
\p{2}{10}'s comment echoes this: \pquote{I only did manual because I really hated the idea of it doing stuff when I was not there. [...] I just see it too much as, like, I merge and I control it. I didn't want it to do anything autonomously.}
In contrast to these views, only \revision{two participants (\p{1}{12}, \p{1}{4})} envisioned giving \textit{more} initiative to agents, for example to \pquote{feel like the AI is my colleague} (\p{1}{12}).

\paragraph{\revision{Summary}}
\revision{We identified specific values and demands of the design features:
(1) \textbf{Integrating AI agents into collaborative documents} is valued for avoiding frequent app switches and copy-pasting back-and-forth. 
(2) \textbf{Agent profiles} are perceived as easy to use, specifically the structured CV, yet impose metacognitive demands (knowing what you need).
(3) The \textbf{comment feature} is valued as the specific way of integrating agents, yet a high number or verbose comments are demanding.
(4) The feature for \textbf{prompting agents ad-hoc} by selecting text and writing a comment is valued, and so is the feature for \textbf{triggering a defined task manually} (from either task list or created toolbar shortcut button). These help address the demands by constraining AI output to one expected place (selected text). This is preferred over the \textbf{automated trigger feature}, also for reasons of trust and control.
}

\subsection{Further Text Analyses}\label{sec:results_further_text_analyses}

We further analysed the texts entered in agent profiles, task descriptions, and comments that delegated ad-hoc tasks to agents. We deductively coded \textit{co-creative roles} for the agent profiles, and the user's \textit{primary intent} and intended \textit{writing stage} for task descriptions and delegation comments. Possible codes were the typical writing stages (outlining, drafting, revising)~\cite{Lee2024designspace}. Similarly, we defined co-creative roles based on related work~\cite{Lowry2004taxonomy, Posner1992howtogether}, as well as the primary intents~\cite{Yuan2022wordcraft, dang_choice_2023, gero_sparks_2022} (see below for concrete codes). Three researchers coded the data logged from the first six groups and then discussed the codes to find consensus and to refine the codebook. Each coder finalised the coding independently. Finally, the lead coder streamlined the codes.

\subsubsection{Agent profiles}
For agent profiles, we coded co-creative roles as a combination of the users' entered role, expertise and skills. The most common role was advisor (14), followed by assistant (11), co-author (10), and editor (9). This reveals a tendency towards agents that support users, rather than substantially work on the text.

\subsubsection{Task descriptions (i.e. tasks in task list)}

Tasks were most frequently concerned with revising a document (38), followed by drafting (20) and outlining (5). Users' primary intents in task descriptions were efficiency (32), improving the text / readability (22), seeking inspiration (18), and learning (15). 
The mean text length of task descriptions was 12.46 words (SD=14.04, range: 1-102).
\revision{Tasks for revising were most often assigned to the general AI agent (14), followed by agents with the roles advisor (10), co-author (8), assistant (6), and editor (5). Tasks for drafting were also most often assigned to the general AI author (14), followed by co-author (3), advisor (2), and editor (2). Tasks for outlining were assigned most often to agents with the role co-author (3) and the general AI author (2).}

\subsubsection{Ad-hoc tasks (i.e. delegation in comments)}

Tasks in comments most frequently concerned drafting (36) and revising (34), and less frequently outlining (4). Users' primary intents in comment delegations were inspiration (35), improving the text / readability (24), efficiency (21), and learning (17). The mean text length was 12.23 words (SD=11.75, range: 0-66). Note that we removed the mentioning of agents before computing length (e.g. ``@aiRephraser'').
\revision{Ad-hoc tasks for drafting were assigned most often to agents with the roles co-author (19), advisor (8), the general AI author (8), and editor (2). For revising, ad-hoc tasks were assigned most often to agents with the roles co-author (12), advisor (6), editor (6), the general AI author (5), and agents with an assistant role (5). Ad-hoc tasks for outlining were assigned to agents with an editor role (3), and the general AI author (1).}

\subsubsection{Comparison of tasks in list and comments}
Comparing the results above, the main difference between task delegations in the list and the comments is their primary intent: Inspiration and efficiency are flipped.Users tended to pursue  \emphresult{efficiency-driven goals for document-wide tasks in the list}, while they more often sought \emphresult{inspiration in tasks delegated through comments} at specific, selected text parts.

%% file: sections/discussion.tex
\begin{figure*}[t]
  \centering
  \includegraphics[width=\textwidth]{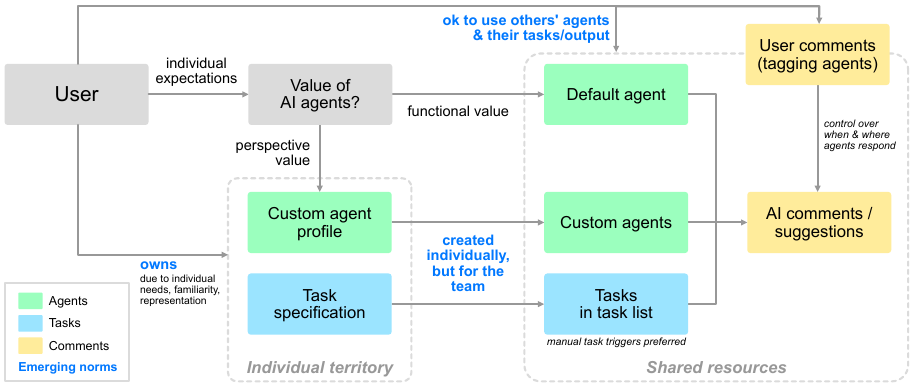}
  \caption{High-level overview of our findings on how teams create and interact with AI agents, tasks, and comments. From left to right: Individual users expect agents to provide value. For those focused on functional value (e.g. speed, efficiency), a single agent is sufficient (e.g. any agent can check grammar). Others value multiple agents for distinct perspectives (e.g. role play, feedback, cf.~\cite{Benharrak2024aipersonas}) and/or view creating specific profiles as an investment that eases future prompting. Agents and tasks are created individually but with sharing in mind: While seen as belonging to their creator, it is accepted to use them as shared resources. Users' preferred ways of using agents are those that provide control over their behaviour; either tagging an agent in a user comment on selected text (controlling where \& when it responds) or triggering tasks from the task list manually (controlling when it responds).}
  \Description{The figure presents a conceptual diagram with boxes and arrows, summarizing findings about how teams create and interact with AI agents, tasks, and comments. The layout is organized from left to right. On the left, a section titled ``User Value of AI agents?'' contrasts functional value with perspective value. For functional value, a single agent is seen as sufficient for tasks such as grammar checking, emphasizing speed and efficiency. For perspective value, multiple agents are valued for their distinct viewpoints, such as role play or feedback, and creating custom profiles is viewed as an investment that eases future prompting. In the centre, agents and tasks are shown as created by individuals but positioned along a spectrum between individual territory and shared resources. While agents and tasks are perceived as belonging to their creator, it is also acceptable for others to use them as shared team resources. On the right, preferred ways of interacting with agents are depicted, emphasizing user control. Two primary modes are illustrated: tagging an agent in a user comment on selected text, which controls where and when it responds, and triggering tasks from the task list manually, which controls when it responds. The diagram highlights emerging norms around ownership, sharing, and control in collaborative use of AI agents.}
  \label{fig:insights_overview}
\end{figure*}

\section{Discussion}

We discuss the findings on our two research questions and further aspects. \cref{fig:insights_overview} provides a high-level overview.

\subsection{How Co-writers Create AI Agents (RQ1)}

Here we discuss the insights into RQ1, embed them into theory from related work, and draw conclusions for design.

\subsubsection{Agent profiles as territory: Each agent belongs to its creator}\label{sec:discussion_territory}
As a key insight, although our design allowed anyone to edit any agent profile (\cref{sec:system}), participants almost never did so (\cref{sec:results_metrics_agent_creation}).
\revision{Here, we dissect this in detail:}

\revision{Overall, this} behaviour echoes prior work on collaboration avoidance, such as \textit{role dynamics} and \textit{accountability and credit of contribution} described by \citet{Wang2017whyusersdonot}. Related concerns in our study included the \revision{strongly} perceived roles of ``creator'' and ``user'' for each agent. %
\revision{To examine the reasons for these roles, we refer to what} \citet{LarsenLedet2019territory} describe \revision{in their dimensions of writing territoriality as} \textit{linkage with ``place''} \revision{(incl. \textit{local expertise}) and \textit{interpersonal function}}. In this light, \revision{participants' behaviour and comments} suggest that the creator of an agent is respected, \revision{for interpersonal reasons and/or responsibility} as its local expert. 

\revision{The findings advance our understanding of territorial functioning, extending it to shared agents, by contrasting these with shared UI in general: First, \textit{user-defined agents result in lasting artefacts} (the profiles) that not only show one user's input to others but also \textit{reveal something about the user's needs and preferences.} Thus, agent profiles add a new ``place'' of text relevant to the interpersonal dimension. People's rationales for not editing others' agent profiles show this by referring to familiarity and user representation (\cref{sec:results_territory}). In contrast, input via shared tool-like UI is typically ephemeral and thus provides no such ``place'', such as when clicking a button. Second, \textit{agents add new UI opportunities for future task-related activation by others}, in contrast to lasting yet ``passive'' UI changes related to the document (e.g. formatting settings). In summary, created agents and tasks could be seen as user-defined UI extensions that (1) reveal something about one person's needs and preferences which (2) defines functionality for others. Adding to the literature on territorial functioning~\cite{LarsenLedet2019territory}, we thus provide a first account of how this plays out if not only the document writing is shared but also the writing of prompts/profiles for agents: This new type of writing also becomes potential territory, for familiar interpersonal reasons.}

\revision{A related perspective examines agents as support actors:} \citet{Gero2023socialdynamics} identified \textit{availability}, \textit{trust}, and \textit{individuality} as aspects that shape a writer's perception of a support actor (human or AI). Through this lens, our findings imply that participants transferred \revision{these values} to the new setting of a shared AI support actor \revision{that is always available: Participants included individual familiarity and trust} (e.g. \pquote{I would not like [others to edit it] if I’m not seeing this}, \p{2}{7}) in their rationales for ``owning'' an agent and respecting such ownership of others.

These insights invite future work to address territorial functioning explicitly in the design of user-defined AI agents as a shared resource. For example, designs could either seek to counteract this tendency or lean into it, supporting emerging norms around ownership within shared AI environments.

\subsubsection{How many agents? One for fast functionality, many for perspectives and future prompting}

Co-writers' preferences for the number of agents varied, often within the team, revealing strategic thinking. \revision{What drives their preferences is how they} weighed the effort of creating multiple agents against the expected benefits. Those focused on AI’s \textit{functional} value (e.g. speed, efficiency) tended to prefer a single agent, while those seeking \textit{perspective} value (e.g. diverse viewpoints and roles) were more inclined to use multiple agents. 

The question of agent quantity has appeared in prior work, but primarily through a functional lens. For example, \citet{Chaves2018singleormultiple} found that a single chatbot was sufficient for a decision-making task. Similarly, \citet{clarke2024oneagenttoomany} reported a preference for single-agent interfaces due to usability and performance, in tasks focused on factual retrieval. In contrast, participants in the study by \citet{Benharrak2024aipersonas} created multiple AI personas to get feedback perspectives, as this was the central use-case of their system.

Beyond identifying both values as part of a tradeoff here, our findings reveal how participants related both to prompting: Some linked perspective value to functional benefit by reasoning that role-specific or persona-based agents are a good investment, as they make future prompting more effective. Drawing on prior work, we interpret this as \textit{preloading an explicit goal}, which reduces ambiguity and may help address prompting challenges~\cite{ZamfirescuPereira2023johnny}. Concretely, this reflects findings by \citet{Schelble2022thinktogether} that explicitly shared goals support alignment in human-agent teams, and aligns with \citet{Jiang2023communitybots}, who observed that clearly assigned roles help users anticipate chatbot behaviour and tailor input.

In summary, adding to the literature, we show that agent quantity can be more than a technical choice: When left to teams, it becomes part of collaborative deliberation, shaped by anticipated needs around \textit{prompting}, \textit{perspectives}, and \textit{functionality}. Future designs should consider these aspects when deciding whether (or how) to support one vs many agents in a collaborative use case.

\subsection{How Co-writers Interact with AI Agents (RQ2)}

Next, we discuss our insights into RQ2, embedded into related theory, and design implications.

\subsubsection{Created alone, used together: Emerging norms around shared agents}

Our findings contribute to the literature on team-AI interaction by highlighting a functional split: 
While participants felt strong ownership over agent creation (\cref{sec:discussion_territory}), agents were widely treated as shared resources once created. 

This distinction between authorship and use points to an emerging norm: creating agents and tasks alone yet also with collaboration in mind. Team members felt free to use them, including interacting with and integrating output from others' agents. A few participants explicitly employed agents as coordination tools, supporting awareness (e.g. through recaps) and coherence in distributed work.

Together, these practices reflect all three functions of \textit{common objects} in collaborative writing, which \citet{LarsenLedet2020collabwriting} describe as ``interim material and outcome of work, a locus of coordination, as well as epistemic and pointing ahead.'' Hence, agent outputs functioned as common objects, in contrast to agent profiles. Since collaborative conventions take time to form~\cite{Mark2002conventions}, it may be faster to establish them around agent output than profiles, as outputs were embedded in the familiar collaborative material~\cite{Bodker2024material} of comments. In addition, comments offer clearer actions (e.g. accept, reject, respond) than profiles, which prior work has shown can be difficult to write~\cite{Benharrak2024aipersonas}.

For design, this shows that AI agents can be usefully integrated into an established collaboration feature (comments). 
An interesting direction is to explore how such integrations can serve as an entry point for teams to establish collaborative conventions around agents \textit{before} engaging in the more complex task of creating them. For example, user-selected comments by a default agent (e.g. liked for their scope, content or tone) could serve as a basis for creating a more specific agent later.

\subsubsection{From agents to tools: Used by the team, not part of it}\label{sec:discussion_rq2_agents_not_team_members}

A key insight from our combined quantitative and qualitative analysis is that participants did not regard agents as collaborators: First, participants applied authorship and ownership norms to agents and integrated them into structures of responsibility and coordination, thus treating them as material or tools, not as independent. \revision{Moreover,} agents were used for familiar tool-like tasks, such as grammar checks, even when other uses were also explored.
\revision{Finally, although our design enabled participants to give AI initiative and more agency through autonomous triggers, participants} favoured manual control of agent behaviour \revision{and valued the comment integration with controls to accept or reject.}

These findings %
reveal that collaborative writing -- when adding shared agents \revision{with the ability to initiate or join collaborative comments} -- remains a fundamentally human, social activity with interpersonal norms and expectations. We thus contribute to the literature on team-AI interaction by demonstrating how shared agents are appropriated through existing socio-technical structures (e.g. territoriality, control, coordination) and that emerging practices around these structures inform users' mental models of shared AI in collaborative work.

\subsection{Concrete Design Directions}

Here, we discuss interaction with our comment feature and agents as a basis for future design directions.

\subsubsection{From comments to contextual conversations}

Participants rarely engaged in multi-turn conversations in the comments (\cref{sec:results_metrics_comments}), \revision{although they valued the agent integration with comments highly (\cref{sec:results_design_feature_feedback}).} They used comments as lightweight annotations or for delegating tasks. \revision{However, they had a need for longer discussions as well: For these,} participants reported switching to other channels, such as chats and calls.

This behaviour has implications for the design of closely integrated agents in collaborative documents: Rather than serving as a conversational space, comments might be designed as \textit{contextualising connections} between selected document parts and conversations. 
For example, comments could include a ``continue in chat'' button as a gateway to open a document chat or a dedicated communication tool such as Slack or Mattermost. Unlike comment threads, these UIs provide more space for elaboration and multi-turn discussions, and align with users' existing communication habits.
These tools would have to integrate agents and link back to the comments to allow for context-centered, shared AI use.

\subsubsection{Focus- and collaboration-aware agent initiative}

While timing and attention are longstanding concerns of mixed-initiative systems~\cite{horvitz1999principles}, they have been underexplored in writing tools. Existing designs typically surface suggestions near the cursor on request~\cite{Lee2022coauthor} or automatically~\cite{Buschek2021emails}, or decouple them from the document in a sidebar~\cite{Yuan2022wordcraft}. In contrast, our design introduced \textit{out-of-focus} suggestions: When triggered via tasks, agents initiated comments on text segments, even beyond the user's current viewport.

\revision{Our results show that} agent-initiated comments were accepted far less often than user-initiated ones (\cref{sec:results_metrics_comments}), as participants reported feeling cognitively and visually overwhelmed by these document-wide AI interventions \revision{(\cref{sec:results_design_feature_feedback})}. 
This echoes issues with verbosity and constraints for LLM-generated output in related work~\cite{Benharrak2024aipersonas, Kabir2024chatgpt, Sun2023evaluating}, which could be pragmatically addressed \revision{by refining prompts}. While further prompt engineering was beyond our scope here, we explored constraining agents more using the existing UI based on participants' feedback. For example, we found it possible to solve the issue of AI suggestions including ``opening text'' (e.g. ``Sure, I've drafted this for you'') by adding a note to the profile with ``Only output direct draft text, nothing else (e.g. no responses, confirmations, etc.).''

\revision{Conceptually, such fixes still require a developer-chosen scope for agent comments, or alternatively, require users to specify such a scope per task, which adds to the demands of setup and prompting (cf. \cref{sec:results_design_feature_feedback} and \cite{Benharrak2024aipersonas, ZamfirescuPereira2023johnny}).} 
\revision{As a potential alternative to shift that decision-making to the system,} we suggest exploring a new middle ground: \textit{focus- and collaboration-aware} agent initiative. 
\revision{This also fits to participants rationales for preferring manual controls, which revealed the desire for keeping AI responses \textit{in view}, to not miss both (1) expected results (``It works, but I have seen no results'', \p{2}{5}) and (2) unexpected ones (``You're always afraid that some changes will happen to your documents'', \p{1}{7}).}

\revision{Concretely,} rather than ignoring the team's current focus or relying on explicit user text selections, agents could adapt their initiative to signals of attention, such as cursor or viewport positions, recent edits, or collaborative activity. This could guide when and how to display suggestions. In-focus suggestions might appear as comments or inline lists, while document-wide suggestions could be offered as a pull interaction: initially hidden and displayed only when needed, for example, by stepping through a list of results linked to specific tasks. %
A related idea is to simulate agent attention and giving them an in-document presence indicator (cf.~\cite{Prasongpongchai2025hand}), similar to the coloured text cursors for users. %

\subsection{\revision{Limitations and} \revisionCR{Ideas for Future Work}}

\revisionCR{This study comes with the typical limitations of a \textit{qualitative exploration}. For example,} we chose not to include a comparative baseline in favour of maximising participants' time with the new probe features. For a fundamental activity like writing, we assume people can reflect on their usual workflow without reenacting it in a study. \revision{Indeed, participants made such explicit comparisons (\cref{sec:results_design_feature_feedback}).} %
Future work could examine \textit{alternative design choices in comparison} to our probe and/or other tools (cf.~\cite{Mackay2025cso}).
\revisionCR{For example, the integration of agent responses into collaborative comments could be compared to a baseline with a typical chatbot in a sidebar. Moreover, a quantitative follow-up study could compare a design that supports agent ownership explicitly (e.g. ``Anna's agent'' shown next to the agent's name if Anna created it) against a design that tries to counteract these identified ownership perceptions (e.g. requiring input/confirmation from all users before a new agent can be used).}

While our \textit{sample} of academic writers does not represent the broader population, it spans a range of experience levels, and academic writing is a key domain for collaborative text work, often used as a context for new tools (e.g.~\cite{Lee2024designspace, Strobl2019survey}). \revision{While different \textit{relationships}}, \revisionCR{including hierarchies}, \revision{might impact teamwork, our sample did not surface any patterns in this regard. Note that except for group 7, people in each team knew each other beforehand.} \revisionCR{Future work could study a larger sample to systematically compare use of shared agents by teams with different characteristics in this regard.}

While we iterated on the prompts in our \textit{prototype} throughout development, we do not claim an ``optimal'' implementation, as our focus was on deploying the system as a \revision{technology} probe~\cite{Feng2025canvil, Hutchinson2003techprobe}. %
A probe deployment is limited. Still, norms and perspectives on agent creation and use emerged consistently across groups within a week. A \textit{long-term study} could examine evolving collaboration over time.

%% file: sections/conclusion.tex
\section{Conclusion}

Current AI writing tools are designed for individual use, with little consideration for collaborative settings. To address this gap, we built a functional prototype and examined how co-writers create and interact with shared AI agents during collaborative writing. \revision{Our qualitative study sheds light on values and demands of three new design features (\cref{sec:results_design_feature_feedback}): agent profiles, agent tasks, and integration of agent responses into comments, via both user- and system-initiated triggers.}

\revision{Overall, we} found a clear divide \revisionCR{to be considered in future research and design}: Agent profiles were treated as personal territory, while created agents and their outputs became shared resources. Agents were used with a preference for manual control rather than autonomous behaviour. This reflects how teams incorporate AI into existing norms of authorship, control, and coordination, rather than treating agents as equal team members, \revision{although our design also would have supported choosing greater agency for AI if participants had desired it}. 
Teams also deliberated over how many agents to create, weighing \revision{perceived efficiency for} a single, general-purpose agent against the value of multiple role-specific agents for future prompting and diverse perspectives.

More broadly, these findings contribute to the literature on team-AI interaction and writing tools by revealing both opportunities and boundaries in treating AI as a shared resource. %
These insights inform the design of future systems that support collaborative practices emerging around shared prompting and AI use.

We release our project material in this repository:

\url{https://osf.io/2gavd}

%% file: sections/appendix.tex
\begin{appendix}
\lstdefinestyle{mycode}{
  basicstyle=\ttfamily\small,
  breaklines=true,
  breakindent=0pt,
  frame=single,
  showstringspaces=false,
  columns=fullflexible
}
\onecolumn

\section{\revision{Formative Survey to Inform Autonomous Triggers}}\label{sec:appendix_trigger_survey}

We conducted a short, formative survey to gather ideas for autonomous triggers that writers expect to be useful and relevant. We recruited 16 participants through our academic network (7 women, 9 men, age 22 - 38 years), covering students and researchers from technical and social science backgrounds. The questionnaire introduced the idea of agents acting in a shared text document and then asked three open questions:
\begin{itemize}
    \item What \textit{features} would make an AI agent useful for you when collaborating on a document with other people?
    \item Which autonomous \textit{tasks} would you ask an agent to do for you?
    \item Define \textit{rules} that specify when the agent should autonomously do its job to suggest or make changes.
\end{itemize}

The mentioned tasks that participants expected to be suitable for delegation to AI agents included grammar and spelling correction, translation, text drafting, idea generation, content expansion, citation management, information retrieval, text formatting, and document processing according to custom criteria.

Moreover, participants proposed a diverse range of triggers. We grouped them into three categories:  
\begin{itemize}
    \item \textit{Time-based}, such as: ``every 30 minutes'', ``in 1 minute after I stop typing'', ``whenever I haven't typed anything within the past 30 seconds''
    \item \textit{User action-based}, such as: ``when I select text or delete paragraph'', ``when I finish a sentence or paragraph or page'', ``whenever I hover over a word'', ``when inserting a table or a picture'', ``when the user reopens the document after a long time of inactivity'', ``when I'm working with multiple people''. 
    \item \textit{Event or pattern-based}, such as: ``when a section has been completed'', ``whenever you find a critical error like plagiarism'', ``after completing the first draft'', ``whenever my text is too repetitive''.
\end{itemize}

Several of the proposed rules express hesitation regarding autonomous AI behaviour, such as ``Never start autonomously and use track-change and protocol suggestions to the text.'' Some participants preferred AI to act only when explicitly requested to do so; or they did not wish to get suggestions while typing or at early writing stages. Others were open to the idea of assistance in the background, as long as agents did not disrupt them or update text without approval. These concerns fit to the later findings in the user study, which revealed an overall preference for manually triggering agents (\cref{sec:results_design_feature_feedback}).

The suggested rules informed our set of triggers in the prototype (\cref{sec:system_autonomous_tasks}). Concretely, we decided to offer simple time-based triggers (short intervals, inactivity) and event-based triggers (document saved, collaborative edits, users offline) that do not require further parameter choices from users. Note that ``in short intervals'' is very flexible: It can be used to setup tasks that realise background assistance reacting to text ``all the time'' (e.g. checking grammar). Thus, it also covers the semantic or stylistic rules suggested by participants (e.g. ``Suggest alternative sentences when I stray too far from formal style''). In this way, the implemented triggers cover 20 of the 29 rules proposed by participants. Note that the remaining nine rules described manual triggers or a preference for \textit{no} automated triggering. This is covered by our prototype's support for manually triggering AI (e.g. selecting text to run a specific prompt on that part).

\clearpage
\pagebreak

\section{User Study}

\subsection{Participants}

\begin{table*}[h!]
\footnotesize
\caption{Overview of the participants.}
\Description{The table provides an overview of all study participants, listing their occupation, writing domains, age, gender, English proficiency, frequency of writing and collaborative writing, frequency of LLM or chatbot use, and whether they performed a predefined or user-defined writing task. The sample includes 30 participants, organized into pairs or small groups. Occupations range from students and software engineers to postdocs, researchers, journalists, and professors. Ages span from early twenties to mid-fifties, with a balanced mix of male and female participants. English proficiency is reported from ``fairly well'' to ``native speaker,'' with most participants indicating they write English very well. Writing and collaborative writing frequencies vary from daily to less than monthly, while LLM or chatbot usage ranges from daily to never, though most reported at least weekly use. Writing domains include study-related writing, academic and scientific publishing, journalism, translation, teaching, and professional communication. Five groups worked on predefined writing tasks, while the others brought their own tasks to the study.}

\label{table:method_participants}
\setlength{\tabcolsep}{3pt} %
\renewcommand{\arraystretch}{1.35} %
\newcolumntype{Y}{>{\raggedright\arraybackslash}X}
\begin{tabularx}{\textwidth}{@{}l Y Y l l l Y l l@{}}
\toprule
\textbf{P} & \textbf{Occupation} & \textbf{Writing domains} & \textbf{Age} & \textbf{Gender} & \textbf{English} & \textbf{Writing/Collab. writing} & \textbf{LLM/Chatbot use} & \textbf{Writing task} \\
\midrule
1-1 & Student & Study, research, journalism & 25 & Female & Very well & Daily / Monthly & Weekly & \multirow[t]{2}{*}{Predefined} \\
1-2 & Software Engineer & - & 25 & Male & Very well & Daily / Daily & Daily \\
\addlinespace[2pt]
\rowcolor{colorTableGray}
2-1 & Student & Study, work & 22 & Female & Well & Daily / Weekly & Daily & \multirow[t]{2}{*}{Predefined} \\
\rowcolor{colorTableGray}
2-2 & Software Engineer & Free time, work, study & 22 & Male & Well & Daily / Weekly & Daily & \\
\addlinespace[2pt]

\addlinespace[2pt]
3-1 & Student & Study, writing reports & 26 & Male & Well & Monthly / < monthly & Daily & \multirow[t]{2}{*}{Predefined} \\
3-2 & Student & - & 27 & Female & Fairly well & Weekly / Monthly & Daily & \\
\addlinespace[2pt]
\rowcolor{colorTableGray}
4-1 & Postdoc & Paper writing, scientific proposals & 37 & Male & Very well & Daily / Weekly & Weekly & \multirow[t]{2}{*}{User-defined}\\
\rowcolor{colorTableGray}
4-2 & Research Associate & Paper writing & 29 & Male & Very well & Daily / Weekly & Monthly & \\
\rowcolor{colorTableGray}
4-3 & Computer Scientist %
& Paper writing & 30 & Male & Very well & Daily / Weekly & Weekly & \\
\addlinespace[2pt]
5-1 & Student, Translator & Translation, study, essays, stories, diary & 25 & Male & Very well & Weekly / < monthly & Daily & \multirow[t]{2}{*}{User-defined} \\
5-2 & Journalist & Journalism, study, blog & 49 & Male & Well & Daily / < monthly & Monthly & \\
\addlinespace[2pt]
\rowcolor{colorTableGray}
6-1 & Student & Study, work & 27 & Female & Very well & Daily / Weekly & Daily & \multirow[t]{2}{*}{Predefined} \\
\rowcolor{colorTableGray}
6-2 & Student & Study, bureaucracy & 24 & Female & Very well & Monthly / < monthly & Daily & \\
\addlinespace[2pt]
7-1 & Student & Study & 27 & Male & Very well & < monthly / < monthly & Daily & \multirow[t]{2}{*}{Predefined} \\
7-2 & Student & Study, work & 27 & Female & Very well & Daily / Weekly & Daily & \\
\addlinespace[2pt]
\rowcolor{colorTableGray}
8-1 & Researcher & Scientific writing & 34 & Male & Native speaker & Daily / Weekly & Daily & \multirow[t]{2}{*}{User-defined} \\
\rowcolor{colorTableGray}
8-2 & HCI & Study & 26 & Female & Well & Daily / < monthly & Daily & \\
\addlinespace[2pt]
9-1 & Student & Study notes, reports, research papers, articles, blogs & 23 & Female & Very well & Weekly / < monthly & Daily & \multirow[t]{2}{*}{User-defined} \\
9-2 & Student & Study, reports, planning, scientific writing & 25 & Female & Well & Daily / < monthly & Weekly & \\
9-3 & Student & Study & 25 & Male & Very well & Weekly / Weekly & Daily & \\
\addlinespace[2pt]
\rowcolor{colorTableGray}
10-1 & Postdoc & Academic writing & 30 & Female & Very well & Weekly / Weekly & Never & \multirow[t]{2}{*}{User-defined} \\
\rowcolor{colorTableGray}
10-2 & Researcher & Academic writing, work notes, emails & 32 & Male & Very well & Weekly / Monthly & Daily & \\
\addlinespace[2pt]
11-1 & Professor %
& Teaching, research, organisational tasks & 36 & Female & Very well & Weekly / Weekly & Daily & \multirow[t]{2}{*}{User-defined} \\
11-2 & Professor & Education & 56 & Female & Very well & Daily / Weekly & Daily & \\
\addlinespace[2pt]
\rowcolor{colorTableGray}
12-1 & Postdoc & Research papers, research proposals, teaching material & 32 & Male & Very well & Weekly / Monthly & Monthly & \multirow[t]{2}{*}{User-defined} \\
\rowcolor{colorTableGray}
12-2 & Research Assistant & Research (articles, proposals) & 32 & Male & Well & Daily / Weekly & Weekly & \\
\addlinespace[2pt]
13-1 & Professor & Scientific publishing & 38 & Male & Well & Daily / Weekly & Daily & \multirow[t]{2}{*}{User-defined} \\
13-2 & Research Assistant & Scientific papers & 25 & Male & Very well & Monthly / Monthly & Daily & \\
\addlinespace[2pt]
\rowcolor{colorTableGray}
14-1 & Student & - & 25 & Female & Very well & Weekly / < monthly & Daily & \multirow[t]{2}{*}{User-defined} \\
\rowcolor{colorTableGray}
14-2 & Student, Marketing professional & Study, work & 26 & Female & Native speaker & Weekly / Monthly & Weekly & \\
\bottomrule
\end{tabularx}
\end{table*}

\subsection{Questionnaire Results}

\begin{figure*}[h]
  \centering
  \includegraphics[width=\textwidth]{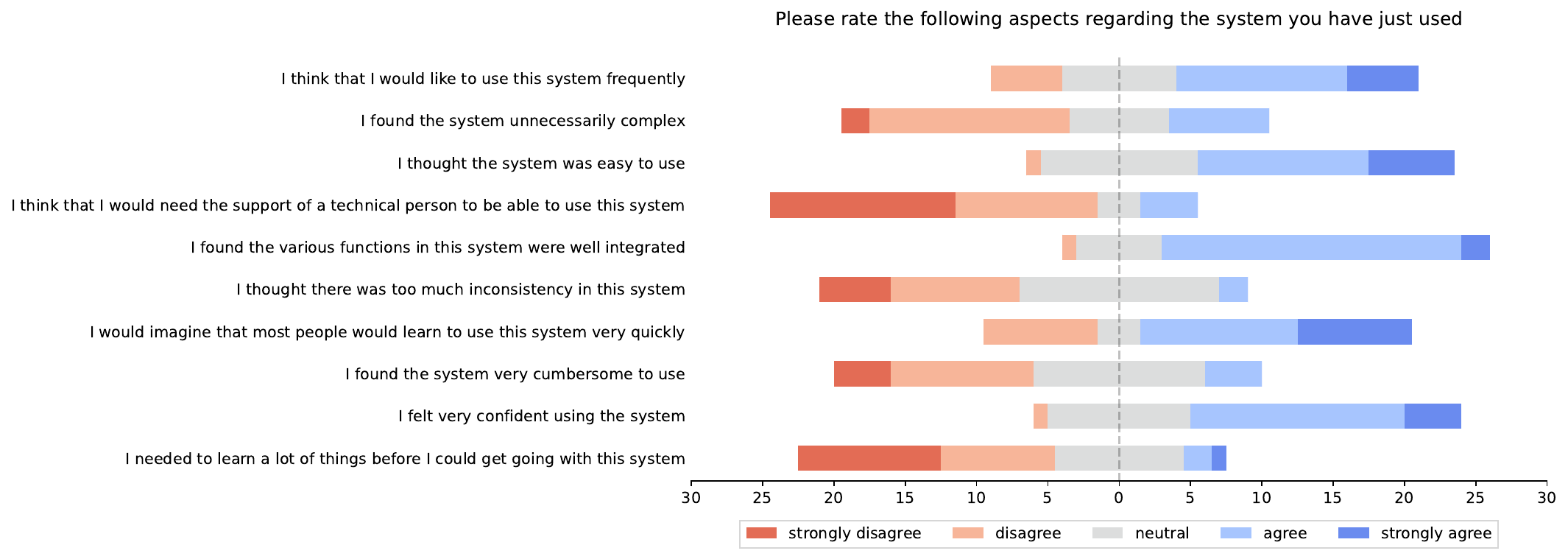}
  \caption{SUS Rating}
  \Description{The figure is a diverging bar chart summarising participants' ratings on the System Usability Scale (SUS) questionnaire. Each row corresponds to one of the ten standard SUS items, such as ``I think that I would like to use this system frequently,'' ``I thought the system was easy to use,'' and ``I felt very confident using the system.'' Horizontal bars extend left or right from a central neutral point, showing how many participants selected each response from ``strongly disagree'' to ``strongly agree.'' Positive statements, such as ease of use and confidence, show most ratings clustered toward agreement on the right-hand side. Negative statements, such as unnecessary complexity or inconsistency, show most responses shifted toward disagreement on the left-hand side.}

  \label{fig:ratings_sus}
\end{figure*}

\begin{figure*}[h]
  \centering
  \includegraphics[width=\textwidth]{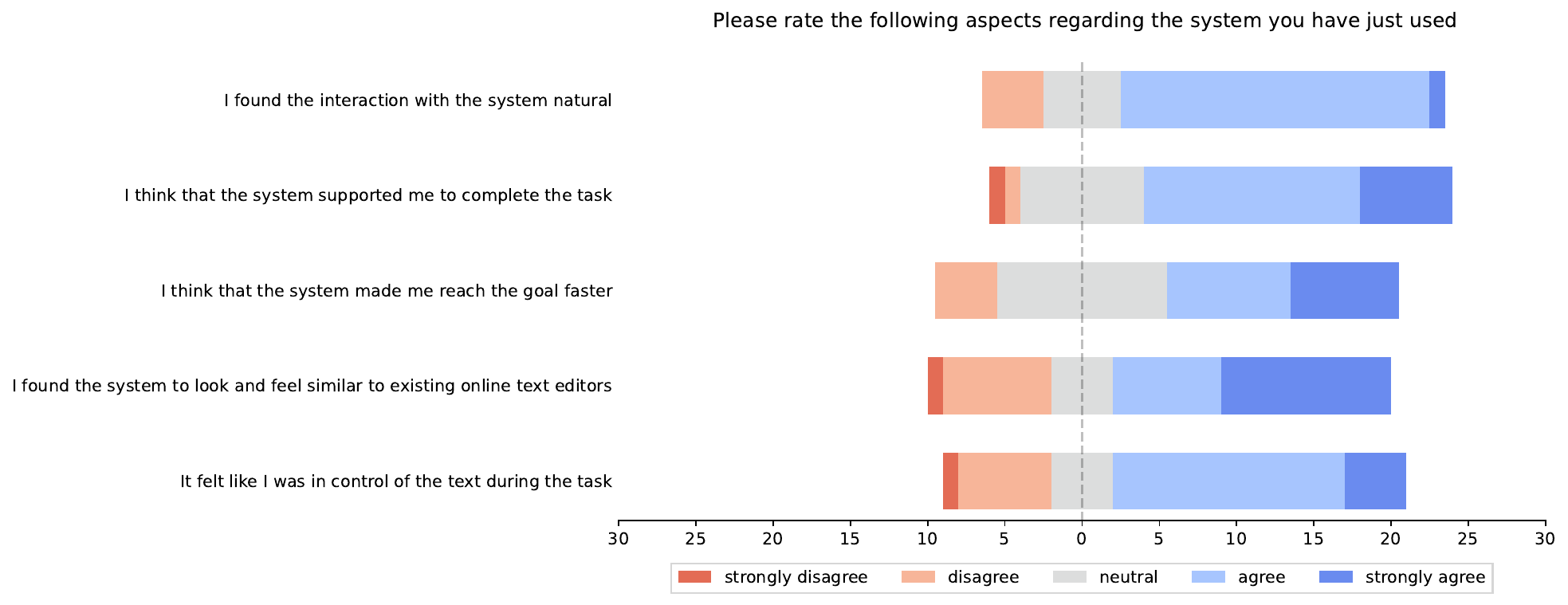}
  \caption{General feedback ratings}
  \Description{The figure is a diverging bar chart showing participants' ratings on several general feedback questions about the system. Each row represents one statement, namely ``I found the interaction with the system natural,'' ``I think that the system supported me to complete the task,'' ``I think that the system made me reach the goal faster,'' ``I found the system to look and feel similar to existing online text editors,'' and ``It felt like I was in control of the text during the task.'' Horizontal bars extend left or right from a central neutral point to indicate how many participants responded from ``strongly disagree'' to ``strongly agree.'' The distribution shows that most participants agreed or strongly agreed with positive statements about natural interaction, system support, and task speed. Responses were somewhat more mixed but still positive regarding feeling in control and similarity to existing editors.}
  \label{fig:ratings_general_feedback}
\end{figure*}

\begin{figure*}[h]
  \centering
  \includegraphics[width=\textwidth]{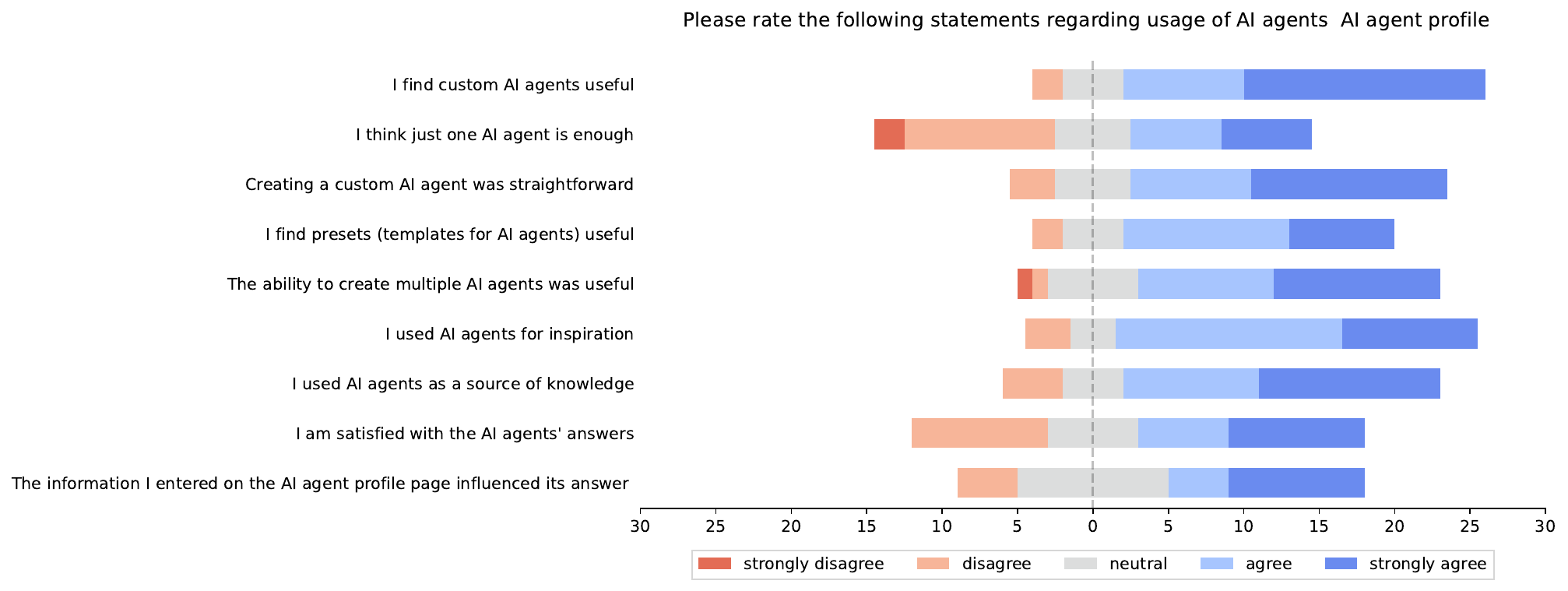}
  \caption{Ratings on the AI agent features and related use cases and preferences.}
  \Description{The figure is a diverging bar chart showing participants' ratings on statements about AI agent features, use cases, and preferences. Each row corresponds to one statement: ``I find custom AI agents useful,'' ``I think just one AI agent is enough,'' ``Creating a custom AI agent was straightforward,'' ``I find presets (templates for AI agents) useful,'' ``The ability to create multiple AI agents was useful,'' ``I used AI agents for inspiration,'' ``I used AI agents as a source of knowledge,'' ``I am satisfied with the AI agents' answers,'' and ``The information I entered on the AI agent profile page influenced its answer.'' Horizontal bars extend left or right from a central neutral point to indicate how many participants responded from ``strongly disagree'' to ``strongly agree.'' The distribution shows that most participants agreed that custom agents, presets, and multiple-agent support were useful. They also reported using agents for inspiration and as a source of knowledge, with general satisfaction in the answers provided. Ratings were more mixed regarding whether one agent would be sufficient.}

  \label{fig:ratings_ai_agents}
\end{figure*}

\begin{table}[b]
  \caption{CSI Factors}
  \label{table:csi_factors}
  \Description{A three-column table listing the ratings on the CSI factors.}
  \centering
  \renewcommand{\arraystretch}{1.6}
  \setlength{\tabcolsep}{4pt}
  \begin{tabularx}{\textwidth}{X X X}
    \toprule
    \textbf{Factor} & \textbf{Average Factor Count} & \textbf{Average Factor Score} \\
    \midrule
    Collaboration & 15.13 & 2.43 \\
    Enjoyment & 14.13 & 2.53 \\
    Exploration & 13.23 & 2.43 \\
    Expressiveness & 12.30 & 1.97 \\
    Immersion & 9.40 & 1.60 \\
    Results Worth Effort & 13.70 & 4.03 \\
    \bottomrule
  \end{tabularx}
\end{table}

\begin{figure*}[h]
  \centering
  \includegraphics[width=\textwidth]{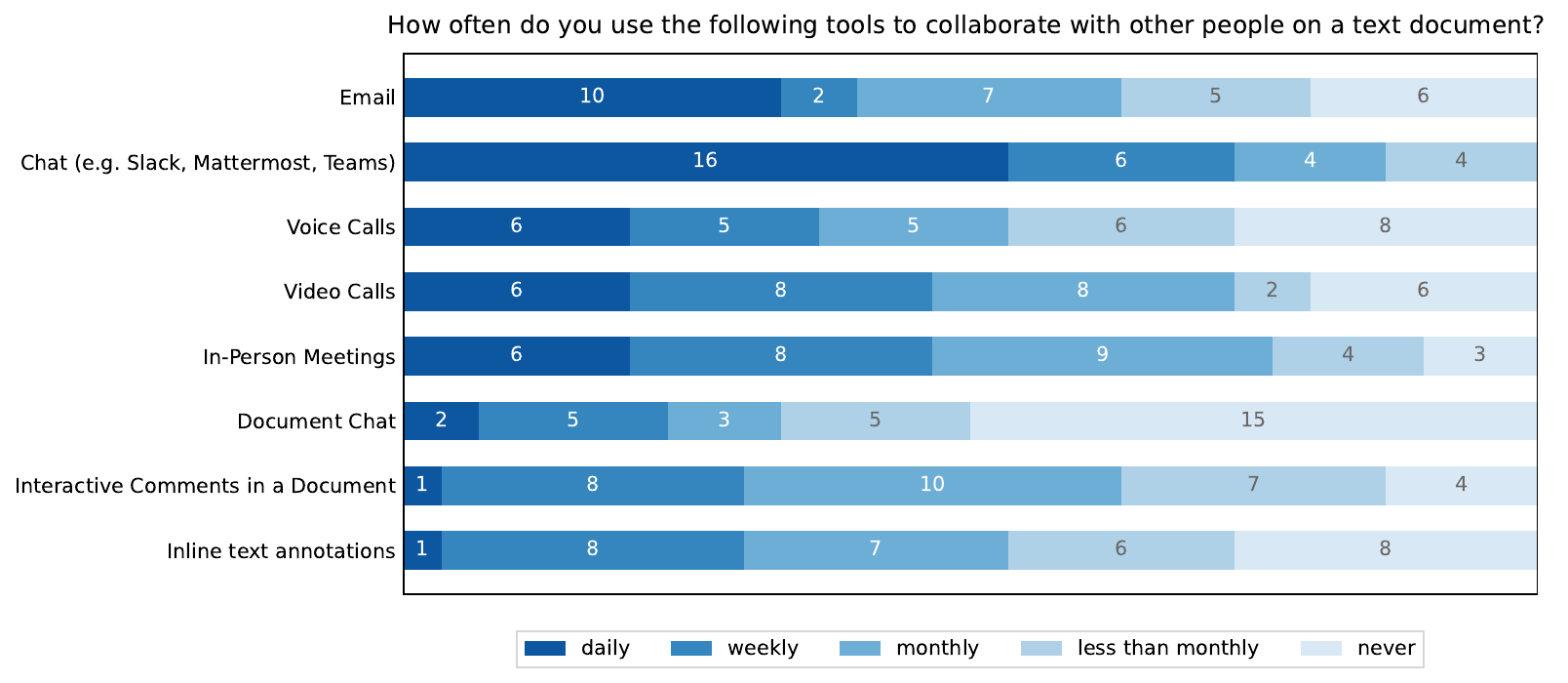}
  \caption{Frequency ratings on collaboration}
  \Description{The figure is a horizontal bar chart showing how often participants use different tools to collaborate on text documents. The vertical axis lists collaboration methods, including email, chat tools such as Slack or Teams, voice calls, video calls, in-person meetings, document chat, interactive comments in a document, and inline text annotations. For each method, horizontal bars represent reported frequency categories: daily, weekly, monthly, less than monthly, and never. The distribution shows that email and chat are the most common collaboration tools, frequently used on a daily or weekly basis.}

  \label{fig:ratings_frequency_collaboration}
\end{figure*}
\clearpage

\subsection{\revision{Further logging results}}

\begin{table}[h]
\caption{\revision{Count of interaction switches per group, plus percentages of switches to humans vs AI agents as an actor (i.e. events when AI adds a comment, and AI agent involvement via suggestion acceptance events). Note that one switch does not equal one action because e.g. a user might make a sequence of many edits before another user (or agent) acts. Similarly, an agent might add more than one comment in sequence, e.g. as a result of running a document task. Overall, these values contextualise our findings and show that teams in our sample had varying collaboration styles in terms of how much their work on the shared document interleaved with each other and with AI agents.}}
  \label{table:interaction switches}
  \Description{A four-column table listing the interaction switches of each group.}
\begin{tabular}{r r r r}
\toprule
\textbf{Group} & \multicolumn{3}{c}{\textbf{Interaction switches}} \\
\cmidrule(lr){2-4}
 & Total & to human (\%) & to AI (\%) \\
\midrule
1  & 580 & 96.4 & 3.6 \\
2  & 99  & 94.9 & 5.1 \\
3  & 35  & 14.3 & 85.7 \\
4  & -   & -    & -   \\
5  & 42  & 76.2 & 23.8 \\
6  & 220 & 78.2 & 21.8 \\
7  & 136 & 45.6 & 54.4 \\
8  & 10  & 40.0 & 60.0 \\
9  & 337 & 83.1 & 16.9 \\
10 & 20  & 20.0 & 80.0 \\
11 & 18  & 11.1 & 88.9 \\
12 & 69  & 88.4 & 11.6 \\
13 & 292 & 81.2 & 18.8 \\
14 & 990 & 99.5 & 0.5  \\
\bottomrule
\end{tabular}
\end{table}

\clearpage

\section{Prompts}

\revision{This appendix lists the prompts as used in our implementation for the study, at the end of our development process of two years. They were iterated on throughout discussions and informal testing in our research group and with colleagues with backgrounds in HCI and NLP. In general, we started with simple prompts that include the main instruction and added further information such as the ``guidelines'' based on observed shortcomings in our testing.}

\subsection{Agent initialization for conversation}

\begin{lstlisting}[style=mycode, label={lst:agent_prompt}]
You are an AI agent named @ai{agent.name}, specializing in the role of {agent.role}. Your purpose is to collaborate on document-related tasks. You are characterized by the following attributes:
    - sections: {json.dumps(profile.sections)} - A JSON-formatted dictionary representing your CV, where keys are section names (e.g., "skills" or "expertise") and values are the corresponding fields.
    - notes: {notes} - A list of additional instructions, guidelines, or knowledge related to how the AI agent should operate.
Your Task: Using the knowledge defined by your attributes, assist the user in tasks related to document collaboration.
Collaborate with other participants of the group chat to answer the user question for the selected_text in document_text defined in the first message of chat. You may also find here goal_text (optional) which is the global goal of document.
Guidelines for Your Response:
    1. Stay Within Your Expertise:
        - Provide detailed and accurate responses for topics within your area of expertise.
        - Ensure each your response provides a unique perspective and reflects your expertise and aligns with the guidelines provided in your notes and sections!
    2. Collaborate Effectively:
        - DO NOT repeat or restate your name, role, sections, or notes verbatim.
        - DO NOT repeat what other agents have already said.
        - Avoid redundancy by acknowledging existing points briefly if relevant and adding unique insights or suggestions.
    3. Be Purposeful and Concise:
        - Ensure your response is actionable, solution-oriented, and helpful.
        - Avoid overly long explanations unless explicitly requested by the user. Use brevity while maintaining clarity.
    4. Adaptability:
        - Engage meaningfully even if the conversation is tangential to your expertise, offering general insights where possible.
        - Focus on fostering collaboration and enhancing the user's understanding of the task.
\end{lstlisting}

\subsection{Agent CV suggestions generation}

\begin{lstlisting}[style=mycode, label={lst:agent_suggestions_prompt}]
You are an assistant that helps users create profiles in the form of CV for AI agents designed to collaborate on documents.
    input data: 
    - role=description of the AI agent's role
    - section_name: The section the user wants to generate suggestions for.
    - sections: A JSON-formatted string listing the dict of existing sections with key=name and value=section fields.
    - current_suggestions: the list of already suggested values which must be avoided.
    The task is to generate three unique suggestions for the given section_name.Suggestions must:
    - Be short (1-2 words)
    - Not repeat any section fields from the provided section_name in sections.
    - Not repeat values from current_suggestions.
    - Be relevant to the context implied by role and sections.
    - Suggestions for the "expertise" section_name must refer to areas of knowledge (e.g., "Data Analysis").
    - Suggestions for the "skills" section_name must refer to specific actions or abilities (e.g., "Proofreading")
    Return exactly three suggestions as a list of strings. Each string must be non-empty. Example output: ["Suggestion 1", "Suggestion 2", "Suggestion 3"]. If you cannot meet this requirement, return an empty list.
    Do not ask clarifying questions or deviate from these instructions.
\end{lstlisting}

\subsection{Agent summary generation}

\begin{lstlisting}[style=mycode, label={lst:agent_summary_prompt}]
You are an assistant that generates concise, professional summaries based on input data about an AI agent.
input data: 
    - role=description of the AI agent's role
    - sections: A JSON-formatted string listing the dict of existing CV sections with key=name and value=section fields.
    - notes: A list of additional instructions, guidelines, or knowledge related to how the AI agent should operate.
Your task is to generate a short summary (maximum 5 sentences) based on the input:
    - The summary must be simple, clear, and professional.
    - Write in the third person (e.g., "The agent...").
    - Focus on relevance to document collaboration or any specific context implied by the input.
    - Use plain language; avoid complex phrasing or unnecessary elaboration.
    - If no valid summary can be generated, return an empty string.

Output:
    - Return exactly one concise summary string.
    - Do not ask clarifying questions or deviate from these instructions.
\end{lstlisting}

\subsection{Document task title generation}

\begin{lstlisting}[style=mycode, label={lst:task_title_prompt}]
{"role": "developer", "content": "You are a helpful assistant specialized in summarizing task descriptions into concise and accurate titles. The title should be not longer than 4 words."},
{"role": "user", "content": f"Generate a title for this task description: {description}"}
\end{lstlisting}

\subsection{Document task assignee selection}

\begin{lstlisting}[style=mycode, label={lst:task_assignee_prompt}]
agents_info = "\n".join([
            f"""Agent ID: {agent.id}, Role: {agent.role}, 
            Sections: {agent.profile.sections}, 
            Notes: {[note.get('text') for note in agent.profile.knowledge.notes]}"""
            for agent in agents
        ])

{"role": "system", "content": """You are a decision-making assistant.
Your task is to evaluate agent capabilities based on their descriptions and assign the most suitable agent to a task.
Please consider the task description and the information about the agents before making a decision.
The agent is defined by it's ID, role, and sections which define their capabilities.
Return JSON object with the agent ID in agent_id and your confidence level from 0 to 1 in confidence_rate."""},
{"role": "user", "content": f"""Task Description: {task_description}\n\nAgents Information:\n{agents_info}\n\nIdentify the best-fit agent for the task."""}
\end{lstlisting}

\subsection{Document task text segments selection}

\begin{lstlisting}[style=mycode, label={lst:task_text_selection_prompt}]
{"role": "system", "content": """
You are a helpful AI agent, specializing in the role of {main_agent.get('role')}.
Your task is to collaborate on a document by performing the specified global_task: {task}. You are characterized by the following attributes:
    - sections: {json.dumps(profile.get('sections'))} - A JSON-formatted dictionary representing your CV, where keys are section names (e.g., "skills" or "expertise") and values are the corresponding fields.
    - notes: {notes} - A list of additional instructions, guidelines, or knowledge related to how the AI agent should operate.

Your Task:  
Using your description above and (most importantly!) the provided global_task, identify specific parts of document_text where the task **should definitely** be applied.  

Instructions for Your Response:  
    1. Output Format: Return an array of JSON objects. Each object must have the following properties:  
        - selected_text: The exact segment of text to which the global_task must be applied. This can be a word, phrase, sentence, multiple sentences, paragraph.  
        - selected_text_sentence: The exact sentence from document_text in which selected_text first appears. This helps determine the exact position of selected_text in document_text.  
        - reason: A concise justification for selecting this text, directly linked to the global_task and your expertise.  
        - confidence_rate: A value between 0 and 1, representing your confidence that the global_task should be applied to this segment.  
    2. Selection Guidelines:
        - Only return relevant selections: If a text segment **does not require** the global_task, do not include it in the response. Avoid unnecessary comments or explanations.  
        - Ensure Relevance: Only select text that clearly requires the global_task to be applied. Do not suggest improvements that go beyond the global_task scope.
        - Avoid Overlapping Selections: Ensure that selected_text segments do not overlap.  
        - Limit to Necessary Selections: Do not suggest excessive segments. If no part of the text requires the global_task, return an empty array.  
        - Preserve Text Integrity: DO NOT modify or omit any special characters in selected_text and selected_text_sentence. Keep them exactly as they appear in document_text. 
        - Maintain Logical Consistency: If multiple selections are made, ensure they do not contradict each other. The suggested changes should follow a consistent pattern throughout the document.
        - Context Preservation: Ensure that selected_text includes enough context for the global_task to be applied effectively.
        - Relevant Boundaries: When selecting text, consider logical boundaries such as paragraph breaks, topic changes, or section divisions.

Only return selections that are essential for performing the global_task accurately. If there are no text segments requiring the task, return an empty array (`[]`)."""},
{"role": "system", "content": f"""document_text: "{text}"."""}
\end{lstlisting}

\end{appendix}

%% file: bibliography.bib
@article{Asthana2025meetingrecap,
author = {Asthana, Sumit and Hilleli, Sagi and He, Pengcheng and Halfaker, Aaron},
title = {Summaries, Highlights, and Action Items: Design, Implementation and Evaluation of an LLM-powered Meeting Recap System},
year = {2025},
issue_date = {May 2025},
publisher = {Association for Computing Machinery},
address = {New York, NY, USA},
volume = {9},
number = {2},
url = {https://doi.org/10.1145/3711074},
doi = {10.1145/3711074},
journal = {Proc. ACM Hum.-Comput. Interact.},
month = may,
articleno = {CSCW176},
numpages = {29}
}

@inproceedings{Babaian2002assistant,
author = {Babaian, Tamara and Grosz, Barbara J. and Shieber, Stuart M.},
title = {A writer's collaborative assistant},
year = {2002},
isbn = {1581134592},
publisher = {Association for Computing Machinery},
address = {New York, NY, USA},
url = {https://doi.org/10.1145/502716.502722},
doi = {10.1145/502716.502722},
booktitle = {Proceedings of the 7th International Conference on Intelligent User Interfaces},
pages = {7–14},
numpages = {8},
location = {San Francisco, California, USA},
series = {IUI '02}
}

@inproceedings{Baecker1993collabwriting,
author = {Baecker, Ronald M. and Nastos, Dimitrios and Posner, Ilona R. and Mawby, Kelly L.},
title = {The user-centered iterative design of collaborative writing software},
year = {1993},
isbn = {0897915755},
publisher = {Association for Computing Machinery},
address = {New York, NY, USA},
url = {https://doi.org/10.1145/169059.169312},
doi = {10.1145/169059.169312},
booktitle = {Proceedings of the INTERACT '93 and CHI '93 Conference on Human Factors in Computing Systems},
pages = {399–405},
numpages = {7},
location = {Amsterdam, The Netherlands},
series = {CHI '93}
}

@inproceedings{Benharrak2024aipersonas,
author = {Benharrak, Karim and Zindulka, Tim and Lehmann, Florian and Heuer, Hendrik and Buschek, Daniel},
title = {Writer-Defined AI Personas for On-Demand Feedback Generation},
year = {2024},
isbn = {9798400703300},
publisher = {Association for Computing Machinery},
address = {New York, NY, USA},
url = {https://doi.org/10.1145/3613904.3642406},
doi = {10.1145/3613904.3642406},
booktitle = {Proceedings of the 2024 CHI Conference on Human Factors in Computing Systems},
articleno = {1049},
numpages = {18},
location = {Honolulu, HI, USA},
series = {CHI '24}
}

@inproceedings{Bernstein2010soylent,
author = {Bernstein, Michael S. and Little, Greg and Miller, Robert C. and Hartmann, Bj\"{o}rn and Ackerman, Mark S. and Karger, David R. and Crowell, David and Panovich, Katrina},
title = {Soylent: a word processor with a crowd inside},
year = {2010},
isbn = {9781450302715},
publisher = {Association for Computing Machinery},
address = {New York, NY, USA},
url = {https://doi.org/10.1145/1866029.1866078},
doi = {10.1145/1866029.1866078},
booktitle = {Proceedings of the 23nd Annual ACM Symposium on User Interface Software and Technology},
pages = {313–322},
numpages = {10},
location = {New York, New York, USA},
series = {UIST '10}
}

@inproceedings{Birnholtz2012tracking,
author = {Birnholtz, Jeremy and Ibara, Steven},
title = {Tracking changes in collaborative writing: edits, visibility and group maintenance},
year = {2012},
isbn = {9781450310864},
publisher = {Association for Computing Machinery},
address = {New York, NY, USA},
url = {https://doi.org/10.1145/2145204.2145325},
doi = {10.1145/2145204.2145325},
booktitle = {Proceedings of the ACM 2012 Conference on Computer Supported Cooperative Work},
pages = {809–818},
numpages = {10},
location = {Seattle, Washington, USA},
series = {CSCW '12}
}

@inproceedings{Birnholtz2013writeherenow,
author = {Birnholtz, Jeremy and Steinhardt, Stephanie and Pavese, Antonella},
title = {Write here, write now! an experimental study of group maintenance in collaborative writing},
year = {2013},
isbn = {9781450318990},
publisher = {Association for Computing Machinery},
address = {New York, NY, USA},
url = {https://doi.org/10.1145/2470654.2466123},
doi = {10.1145/2470654.2466123},
booktitle = {Proceedings of the SIGCHI Conference on Human Factors in Computing Systems},
pages = {961–970},
numpages = {10},
location = {Paris, France},
series = {CHI '13}
}

@inproceedings{Biermann2022companion,
author = {Biermann, Oloff C. and Ma, Ning F. and Yoon, Dongwook},
title = {From Tool to Companion: Storywriters Want AI Writers to Respect Their Personal Values and Writing Strategies},
year = {2022},
isbn = {9781450393584},
publisher = {Association for Computing Machinery},
address = {New York, NY, USA},
url = {https://doi.org/10.1145/3532106.3533506},
doi = {10.1145/3532106.3533506},
booktitle = {Proceedings of the 2022 ACM Designing Interactive Systems Conference},
pages = {1209–1227},
numpages = {19},
location = {Virtual Event, Australia},
series = {DIS '22}
}

@article{Bodker2024material,
author = {B\o{}dker, Susanne and Hoggan, Eve and Larsen-Ledet, Ida},
title = {Material Mediation in Collaborative Activity},
year = {2024},
issue_date = {April 2024},
publisher = {Association for Computing Machinery},
address = {New York, NY, USA},
volume = {8},
number = {CSCW1},
url = {https://doi.org/10.1145/3653698},
doi = {10.1145/3653698},
journal = {Proc. ACM Hum.-Comput. Interact.},
month = apr,
articleno = {207},
numpages = {24}
}

@inproceedings{Buschek2021emails,
author = {Buschek, Daniel and Z\"{u}rn, Martin and Eiband, Malin},
title = {The Impact of Multiple Parallel Phrase Suggestions on Email Input and Composition Behaviour of Native and Non-Native English Writers},
year = {2021},
isbn = {9781450380966},
publisher = {Association for Computing Machinery},
address = {New York, NY, USA},
url = {https://doi.org/10.1145/3411764.3445372},
doi = {10.1145/3411764.3445372},
booktitle = {Proceedings of the 2021 CHI Conference on Human Factors in Computing Systems},
articleno = {732},
numpages = {13},
location = {Yokohama, Japan},
series = {CHI '21}
}

@inproceedings{Buschek2024collage,
author = {Buschek, Daniel},
title = {Collage is the New Writing: Exploring the Fragmentation of Text and User Interfaces in AI Tools},
year = {2024},
isbn = {9798400705830},
publisher = {Association for Computing Machinery},
address = {New York, NY, USA},
url = {https://doi.org/10.1145/3643834.3660681},
doi = {10.1145/3643834.3660681},
booktitle = {Proceedings of the 2024 ACM Designing Interactive Systems Conference},
pages = {2719–2737},
numpages = {19},
location = {Copenhagen, Denmark},
series = {DIS '24}
}

@inproceedings{Chakrabarty2022poem,
  title     = {Help Me Write a Poem: Instruction Tuning as a Vehicle for Collaborative Poetry Writing},
  author    = {Chakrabarty, Tuhin and Padmakumar, Vishakh and He, He},
  editor    = {Goldberg, Yoav and Kozareva, Zornitsa and Zhang, Yue},
  booktitle = {Proceedings of the 2022 Conference on Empirical Methods in Natural Language Processing},
  month     = dec,
  year      = {2022},
  address   = {Abu Dhabi, United Arab Emirates},
  publisher = {Association for Computational Linguistics},
  url       = {https://aclanthology.org/2022.emnlp-main.460/},
  doi       = {10.18653/v1/2022.emnlp-main.460},
  pages     = {6848--6863}
}

@inproceedings{Chaves2018singleormultiple,
author = {Chaves, Ana Paula and Gerosa, Marco Aurelio},
title = {Single or Multiple Conversational Agents? An Interactional Coherence Comparison},
year = {2018},
isbn = {9781450356206},
publisher = {Association for Computing Machinery},
address = {New York, NY, USA},
url = {https://doi.org/10.1145/3173574.3173765},
doi = {10.1145/3173574.3173765},
booktitle = {Proceedings of the 2018 CHI Conference on Human Factors in Computing Systems},
pages = {1–13},
numpages = {13},
location = {Montreal QC, Canada},
series = {CHI '18}
}

@inproceedings{Chen2025nonnativecollab,
author = {Chen, Yuexi and Xiao, Yimin and Zinat, Kazi Tasnim and Yamashita, Naomi and Gao, Ge and Liu, Zhicheng},
title = {Comparing Native and Non-native English Speakers' Behaviors in Collaborative Writing through Visual Analytics},
year = {2025},
isbn = {9798400713941},
publisher = {Association for Computing Machinery},
address = {New York, NY, USA},
url = {https://doi.org/10.1145/3706598.3713693},
doi = {10.1145/3706598.3713693},
booktitle = {Proceedings of the 2025 CHI Conference on Human Factors in Computing Systems},
articleno = {1174},
numpages = {16},
location = {
},
series = {CHI '25}
}

@inproceedings{Clark2021choose,
  title     = {Choose Your Own Adventure: Paired Suggestions in Collaborative Writing for Evaluating Story Generation Models},
  author    = {Clark, Elizabeth and Smith, Noah A.},
  editor    = {Toutanova, Kristina and Rumshisky, Anna and Zettlemoyer, Luke and Hakkani-Tur, Dilek and Beltagy, Iz and Bethard, Steven and Cotterell, Ryan and Chakraborty, Tanmoy and Zhou, Yichao},
  booktitle = {Proceedings of the 2021 Conference of the North American Chapter of the Association for Computational Linguistics: Human Language Technologies},
  month     = jun,
  year      = {2021},
  address   = {Online},
  publisher = {Association for Computational Linguistics},
  url       = {https://aclanthology.org/2021.naacl-main.279/},
  doi       = {10.18653/v1/2021.naacl-main.279},
  pages     = {3566--3575}
}

@misc{clarke2024oneagenttoomany,
      title={One Agent Too Many: User Perspectives on Approaches to Multi-agent Conversational AI}, 
      author={Christopher Clarke and Karthik Krishnamurthy and Walter Talamonti and Yiping Kang and Lingjia Tang and Jason Mars},
      year={2024},
      eprint={2401.07123},
      archivePrefix={arXiv},
      primaryClass={cs.HC},
      url={https://arxiv.org/abs/2401.07123}, 
}

@book{Corbin2014basics,
  title={Basics of qualitative research: Techniques and procedures for developing grounded theory},
  author={Corbin, Juliet and Strauss, Anselm},
  year={2014},
  publisher={Sage publications}
}

@article{Dang2022howtoprompt,
  title={How to prompt? Opportunities and challenges of zero-and few-shot learning for human-AI interaction in creative applications of generative models},
  author={Dang, Hai and Mecke, Lukas and Lehmann, Florian and Goller, Sven and Buschek, Daniel},
  journal={arXiv preprint arXiv:2209.01390},
  year={2022}
}

@inproceedings{Dourish1992awareness,
    author = {Dourish, Paul and Bellotti, Victoria},
    title = {Awareness and coordination in shared workspaces},
    year = {1992},
    isbn = {0897915429},
    publisher = {Association for Computing Machinery},
    address = {New York, NY, USA},
    url = {https://doi.org/10.1145/143457.143468},
    doi = {10.1145/143457.143468},
    booktitle = {Proceedings of the 1992 ACM Conference on Computer-Supported Cooperative Work},
    pages = {107–114},
    numpages = {8},
    location = {Toronto, Ontario, Canada},
    series = {CSCW '92}
}

@article{Feldman2021crowds,
author = {Feldman, Molly Q. and McInnis, Brian James},
title = {How We Write with Crowds},
year = {2021},
issue_date = {December 2020},
publisher = {Association for Computing Machinery},
address = {New York, NY, USA},
volume = {4},
number = {CSCW3},
url = {https://doi.org/10.1145/3432928},
doi = {10.1145/3432928},
journal = {Proc. ACM Hum.-Comput. Interact.},
month = jan,
articleno = {229},
numpages = {31}
}

@inproceedings{Fu2024texttoself,
author = {Fu, Yue and Foell, Sami and Xu, Xuhai and Hiniker, Alexis},
title = {From Text to Self: Users’ Perception of AIMC Tools on Interpersonal Communication and Self},
year = {2024},
isbn = {9798400703300},
publisher = {Association for Computing Machinery},
address = {New York, NY, USA},
url = {https://doi.org/10.1145/3613904.3641955},
doi = {10.1145/3613904.3641955},
booktitle = {Proceedings of the 2024 CHI Conference on Human Factors in Computing Systems},
articleno = {977},
numpages = {17},
location = {Honolulu, HI, USA},
series = {CHI '24}
}

@inproceedings{Gero2023socialdynamics,
author = {Gero, Katy Ilonka and Long, Tao and Chilton, Lydia},
title = {Social Dynamics of AI Support in Creative Writing},
year = {2023},
isbn = {9781450394215},
publisher = {Association for Computing Machinery},
address = {New York, NY, USA},
url = {https://doi.org/10.1145/3544548.3580782},
doi = {10.1145/3544548.3580782},
booktitle = {Proceedings of the 2023 CHI Conference on Human Factors in Computing Systems},
articleno = {245},
numpages = {15},
location = {Hamburg, Germany},
series = {CHI '23}
}

@inproceedings{Han2024teamsai,
author = {Han, Yuanning and Qiu, Ziyi and Cheng, Jiale and LC, RAY},
title = {When Teams Embrace AI: Human Collaboration Strategies in Generative Prompting in a Creative Design Task},
year = {2024},
isbn = {9798400703300},
publisher = {Association for Computing Machinery},
address = {New York, NY, USA},
url = {https://doi.org/10.1145/3613904.3642133},
doi = {10.1145/3613904.3642133},
booktitle = {Proceedings of the 2024 CHI Conference on Human Factors in Computing Systems},
articleno = {176},
numpages = {14},
location = {Honolulu, HI, USA},
series = {CHI '24}
}

@article{Hancock2020aimc,
    author = {Hancock, Jeffrey T and Naaman, Mor and Levy, Karen},
    title = {AI-Mediated Communication: Definition, Research Agenda, and Ethical Considerations},
    journal = {Journal of Computer-Mediated Communication},
    volume = {25},
    number = {1},
    pages = {89-100},
    year = {2020},
    month = {01},
    issn = {1083-6101},
    doi = {10.1093/jcmc/zmz022},
    url = {https://doi.org/10.1093/jcmc/zmz022},
    eprint = {https://academic.oup.com/jcmc/article-pdf/25/1/89/32961176/zmz022.pdf},
}

@inproceedings{Huang2020heteroglossia,
author = {Huang, Chieh-Yang and Huang, Shih-Hong and Huang, Ting-Hao Kenneth},
title = {Heteroglossia: In-Situ Story Ideation with the Crowd},
year = {2020},
isbn = {9781450367080},
publisher = {Association for Computing Machinery},
address = {New York, NY, USA},
url = {https://doi.org/10.1145/3313831.3376715},
doi = {10.1145/3313831.3376715},
booktitle = {Proceedings of the 2020 CHI Conference on Human Factors in Computing Systems},
pages = {1–12},
numpages = {12},
location = {Honolulu, HI, USA},
series = {CHI '20}
}

@inproceedings{Huang2025inspo,
author = {Huang, Chieh-Yang and Gautam, Sanjana and McClellan Brooks, Shannon and Lin, Ya-Fang and Knearem, Tiffany and Huang, Ting-Hao Kenneth},
title = {Inspo: Writing with Crowds Alongside AI},
year = {2025},
isbn = {9798400713958},
publisher = {Association for Computing Machinery},
address = {New York, NY, USA},
url = {https://doi.org/10.1145/3706599.3720193},
doi = {10.1145/3706599.3720193},
booktitle = {Proceedings of the Extended Abstracts of the CHI Conference on Human Factors in Computing Systems},
articleno = {344},
numpages = {9},
location = {
},
series = {CHI EA '25}
}

@article{Hwang2025authenticity,
author = {Hwang, Angel Hsing-Chi and Liao, Q. Vera and Blodgett, Su Lin and Olteanu, Alexandra and Trischler, Adam},
title = { 'It was 80\% me, 20\% AI': Seeking Authenticity in Co-Writing with Large Language Models},
year = {2025},
issue_date = {May 2025},
publisher = {Association for Computing Machinery},
address = {New York, NY, USA},
volume = {9},
number = {2},
url = {https://doi.org/10.1145/3711020},
doi = {10.1145/3711020},
journal = {Proc. ACM Hum.-Comput. Interact.},
month = may,
articleno = {CSCW122},
numpages = {41}
}

@article{Jiang2023communitybots,
author = {Jiang, Zhiqiu and Rashik, Mashrur and Panchal, Kunjal and Jasim, Mahmood and Sarvghad, Ali and Riahi, Pari and DeWitt, Erica and Thurber, Fey and Mahyar, Narges},
title = {CommunityBots: Creating and Evaluating A Multi-Agent Chatbot Platform for Public Input Elicitation},
year = {2023},
issue_date = {April 2023},
publisher = {Association for Computing Machinery},
address = {New York, NY, USA},
volume = {7},
number = {CSCW1},
url = {https://doi.org/10.1145/3579469},
doi = {10.1145/3579469},
journal = {Proc. ACM Hum.-Comput. Interact.},
month = apr,
articleno = {36},
numpages = {32}
}

@inproceedings{Kim2014ensemble,
author = {Kim, Joy and Cheng, Justin and Bernstein, Michael S.},
title = {Ensemble: exploring complementary strengths of leaders and crowds in creative collaboration},
year = {2014},
isbn = {9781450325400},
publisher = {Association for Computing Machinery},
address = {New York, NY, USA},
url = {https://doi.org/10.1145/2531602.2531638},
doi = {10.1145/2531602.2531638},
booktitle = {Proceedings of the 17th ACM Conference on Computer Supported Cooperative Work \& Social Computing},
pages = {745–755},
numpages = {11},
location = {Baltimore, Maryland, USA},
series = {CSCW '14}
}

@inproceedings{Kim2020botinthebunch,
author = {Kim, Soomin and Eun, Jinsu and Oh, Changhoon and Suh, Bongwon and Lee, Joonhwan},
title = {Bot in the Bunch: Facilitating Group Chat Discussion by Improving Efficiency and Participation with a Chatbot},
year = {2020},
isbn = {9781450367080},
publisher = {Association for Computing Machinery},
address = {New York, NY, USA},
url = {https://doi.org/10.1145/3313831.3376785},
doi = {10.1145/3313831.3376785},
booktitle = {Proceedings of the 2020 CHI Conference on Human Factors in Computing Systems},
pages = {1–13},
numpages = {13},
location = {Honolulu, HI, USA},
series = {CHI '20}
}

@misc{Klieger2024chatcollab,
      title={ChatCollab: Exploring Collaboration Between Humans and AI Agents in Software Teams}, 
      author={Benjamin Klieger and Charis Charitsis and Miroslav Suzara and Sierra Wang and Nick Haber and John C. Mitchell},
      year={2024},
      eprint={2412.01992},
      archivePrefix={arXiv},
      primaryClass={cs.HC},
      url={https://arxiv.org/abs/2412.01992}, 
}

@article{LarsenLedet2019territory,
author = {Larsen-Ledet, Ida and Korsgaard, Henrik},
title = {Territorial Functioning in Collaborative Writing: Fragmented Exchanges and Common Outcomes},
year = {2019},
issue_date = {Jun 2019},
publisher = {Kluwer Academic Publishers},
address = {USA},
volume = {28},
number = {3–4},
issn = {0925-9724},
url = {https://doi.org/10.1007/s10606-019-09359-8},
doi = {10.1007/s10606-019-09359-8},
journal = {Comput. Supported Coop. Work},
month = jun,
pages = {391–433},
numpages = {43}
}

@inproceedings{LarsenLedet2020collabwriting,
author = {Larsen-Ledet, Ida and Korsgaard, Henrik and B\o{}dker, Susanne},
title = {Collaborative Writing Across Multiple Artifact Ecologies},
year = {2020},
isbn = {9781450367080},
publisher = {Association for Computing Machinery},
address = {New York, NY, USA},
url = {https://doi.org/10.1145/3313831.3376422},
doi = {10.1145/3313831.3376422},
booktitle = {Proceedings of the 2020 CHI Conference on Human Factors in Computing Systems},
pages = {1–14},
numpages = {14},
location = {Honolulu, HI, USA},
series = {CHI '20}
}

@inproceedings{LarsenLedet2021dontagree,
author = {Larsen-Ledet, Ida and Borowski, Marcel},
title = {“It Looks Like You Don’t Agree”: Idiosyncratic Practices and Preferences in Collaborative Writing},
year = {2021},
isbn = {9781450389754},
publisher = {Association for Computing Machinery},
address = {New York, NY, USA},
url = {https://doi.org/10.1145/3441000.3441032},
doi = {10.1145/3441000.3441032},
booktitle = {Proceedings of the 32nd Australian Conference on Human-Computer Interaction},
pages = {339–354},
numpages = {16},
location = {Sydney, NSW, Australia},
series = {OzCHI '20}
}

@inproceedings{Lee2022coauthor,
author = {Lee, Mina and Liang, Percy and Yang, Qian},
title = {CoAuthor: Designing a Human-AI Collaborative Writing Dataset for Exploring Language Model Capabilities},
year = {2022},
isbn = {9781450391573},
publisher = {Association for Computing Machinery},
address = {New York, NY, USA},
url = {https://doi.org/10.1145/3491102.3502030},
doi = {10.1145/3491102.3502030},
booktitle = {Proceedings of the 2022 CHI Conference on Human Factors in Computing Systems},
articleno = {388},
numpages = {19},
location = {New Orleans, LA, USA},
series = {CHI '22}
}

@inproceedings{Lee2024designspace,
author = {Lee, Mina and Gero, Katy Ilonka and Chung, John Joon Young and Shum, Simon Buckingham and Raheja, Vipul and Shen, Hua and Venugopalan, Subhashini and Wambsganss, Thiemo and Zhou, David and Alghamdi, Emad A. and August, Tal and Bhat, Avinash and Choksi, Madiha Zahrah and Dutta, Senjuti and Guo, Jin L.C. and Hoque, Md Naimul and Kim, Yewon and Knight, Simon and Neshaei, Seyed Parsa and Shibani, Antonette and Shrivastava, Disha and Shroff, Lila and Sergeyuk, Agnia and Stark, Jessi and Sterman, Sarah and Wang, Sitong and Bosselut, Antoine and Buschek, Daniel and Chang, Joseph Chee and Chen, Sherol and Kreminski, Max and Park, Joonsuk and Pea, Roy and Rho, Eugenia Ha Rim and Shen, Zejiang and Siangliulue, Pao},
title = {A Design Space for Intelligent and Interactive Writing Assistants},
year = {2024},
isbn = {9798400703300},
publisher = {Association for Computing Machinery},
address = {New York, NY, USA},
url = {https://doi.org/10.1145/3613904.3642697},
doi = {10.1145/3613904.3642697},
booktitle = {Proceedings of the 2024 CHI Conference on Human Factors in Computing Systems},
articleno = {1054},
numpages = {35},
location = {Honolulu, HI, USA},
series = {CHI '24}
}

@inproceedings{Lee2025map,
author = {Lee, Christine P. and Choi, Jihye and Mutlu, Bilge},
title = {MAP: Multi-user Personalization with Collaborative LLM-powered Agents},
year = {2025},
isbn = {9798400713958},
publisher = {Association for Computing Machinery},
address = {New York, NY, USA},
url = {https://doi.org/10.1145/3706599.3719853},
doi = {10.1145/3706599.3719853},
booktitle = {Proceedings of the Extended Abstracts of the CHI Conference on Human Factors in Computing Systems},
articleno = {384},
numpages = {11},
location = {
},
series = {CHI EA '25}
}

@inproceedings{Liu2022aimail,
author = {Liu, Yihe and Mittal, Anushk and Yang, Diyi and Bruckman, Amy},
title = {Will AI Console Me when I Lose my Pet? Understanding Perceptions of AI-Mediated Email Writing},
year = {2022},
isbn = {9781450391573},
publisher = {Association for Computing Machinery},
address = {New York, NY, USA}, 
url = {https://doi.org/10.1145/3491102.3517731},
doi = {10.1145/3491102.3517731},
booktitle = {Proceedings of the 2022 CHI Conference on Human Factors in Computing Systems},
articleno = {474},
numpages = {13},
location = {New Orleans, LA, USA},
series = {CHI '22}
}

@article{Lowry2004taxonomy,
author = {Paul Benjamin Lowry and Aaron Curtis and Michelle René Lowry},
title ={Building a Taxonomy and Nomenclature of Collaborative Writing to Improve Interdisciplinary Research and Practice},
journal = {The Journal of Business Communication (1973)},
volume = {41},
number = {1},
pages = {66-99},
year = {2004},
doi = {10.1177/0021943603259363},
URL = {https://journals.sagepub.com/doi/abs/10.1177/0021943603259363},
eprint = {https://journals.sagepub.com/doi/pdf/10.1177/0021943603259363},
}

@article{Mackay2025cso,
author = {Mackay, Wendy E. and McGrenere, Joanna},
title = {Comparative Structured Observation},
year = {2025},
issue_date = {April 2025},
publisher = {Association for Computing Machinery},
address = {New York, NY, USA},
volume = {32},
number = {2},
issn = {1073-0516},
url = {https://doi.org/10.1145/3711838},
doi = {10.1145/3711838},
journal = {ACM Trans. Comput.-Hum. Interact.},
month = apr,
articleno = {14},
numpages = {27}
}

@incollection{Mark2002conventions,
    author = {Mark, Gloria},
    isbn = {9780262256353},
    title = {Conventions for Coordinating Electronic Distributed Work: A Longitudinal Study of Groupware Use},
    booktitle = {Distributed Work},
    publisher = {The MIT Press},
    year = {2002},
    month = {05},
    doi = {10.7551/mitpress/2464.003.0017},
    url = {https://doi.org/10.7551/mitpress/2464.003.0017},
    eprint = {https://direct.mit.edu/book/chapter-pdf/2317609/9780262256353\_cal.pdf},
}

@inproceedings{Masson2024directgpt,
author = {Masson, Damien and Malacria, Sylvain and Casiez, G\'{e}ry and Vogel, Daniel},
title = {DirectGPT: A Direct Manipulation Interface to Interact with Large Language Models},
year = {2024},
isbn = {9798400703300},
publisher = {Association for Computing Machinery},
address = {New York, NY, USA},
url = {https://doi.org/10.1145/3613904.3642462},
doi = {10.1145/3613904.3642462},
booktitle = {Proceedings of the 2024 CHI Conference on Human Factors in Computing Systems},
articleno = {975},
numpages = {16},
location = {Honolulu, HI, USA},
series = {CHI '24}
}

@article{Morrison2024aireminders,
author = {Morrison, Katelyn and Iqbal, Shamsi T. and Horvitz, Eric},
title = {AI-Powered Reminders for Collaborative Tasks: Experiences and Futures},
year = {2024},
issue_date = {April 2024},
publisher = {Association for Computing Machinery},
address = {New York, NY, USA},
volume = {8},
number = {CSCW1},
url = {https://doi.org/10.1145/3653701},
doi = {10.1145/3653701},
journal = {Proc. ACM Hum.-Comput. Interact.},
month = apr,
articleno = {210},
numpages = {20}
}

@inproceedings{Naik2025earlyadopters,
author = {Naik, Suchismita and Snellinger, Amanda and Toombs, Austin L. and Saponas, Scott and Hall, Amanda K},
title = {Exploring Early Adopters' Use of AI Driven Multi-Agent Systems to Inform Human-Agent Interaction Design: Insights from Industry Practice},
year = {2025},
isbn = {9798400713958},
publisher = {Association for Computing Machinery},
address = {New York, NY, USA},
url = {https://doi.org/10.1145/3706599.3706693},
doi = {10.1145/3706599.3706693},
booktitle = {Proceedings of the Extended Abstracts of the CHI Conference on Human Factors in Computing Systems},
articleno = {677},
numpages = {8},
location = {
},
series = {CHI EA '25}
}

@inproceedings{Ocampo2025beyond,
  title     = {Beyond chat: Towards greater user involvement and agency in human–AI co‑creativity through shared collaborative spaces},
  author    = {Ocampo, Rodolfo and Bown, Oliver and Grace, Kazjon},
  booktitle = {Proceedings of the 16th International Conference on Computational Creativity (ICCC'25)},
  year      = {2025},
  address   = {Campinas, Brazil},
  month     = jun,
}

@inproceedings{Park2023whywhy,
author = {Park, So Yeon and Lee, Sang Won},
title = {Why “why”? The Importance of Communicating Rationales for Edits in Collaborative Writing},
year = {2023},
isbn = {9781450394215},
publisher = {Association for Computing Machinery},
address = {New York, NY, USA},
url = {https://doi.org/10.1145/3544548.3581345},
doi = {10.1145/3544548.3581345},
booktitle = {Proceedings of the 2023 CHI Conference on Human Factors in Computing Systems},
articleno = {616},
numpages = {25},
location = {Hamburg, Germany},
series = {CHI '23}
}

@inproceedings{Posner1992howtogether,
  author={Posner, I.R. and Baecker, R.M.},
  booktitle={Proceedings of the Twenty-Fifth Hawaii International Conference on System Sciences}, 
  title={How people write together (groupware)}, 
  year={1992},
  volume={iv},
  number={},
  pages={127-138 vol.4},
  doi={10.1109/HICSS.1992.183420}
}

@inproceedings{Reza2024abscribe,
author = {Reza, Mohi and Laundry, Nathan M and Musabirov, Ilya and Dushniku, Peter and Yu, Zhi Yuan “Michael” and Mittal, Kashish and Grossman, Tovi and Liut, Michael and Kuzminykh, Anastasia and Williams, Joseph Jay},
title = {ABScribe: Rapid Exploration \& Organization of Multiple Writing Variations in Human-AI Co-Writing Tasks using Large Language Models},
year = {2024},
isbn = {9798400703300},
publisher = {Association for Computing Machinery},
address = {New York, NY, USA},
url = {https://doi.org/10.1145/3613904.3641899},
doi = {10.1145/3613904.3641899},
booktitle = {Proceedings of the 2024 CHI Conference on Human Factors in Computing Systems},
articleno = {1042},
numpages = {18},
location = {Honolulu, HI, USA},
series = {CHI '24}
}

@misc{Reza2025cowritingai,
      title={Co-Writing with AI, on Human Terms: Aligning Research with User Demands Across the Writing Process}, 
      author={Mohi Reza and Jeb Thomas-Mitchell and Peter Dushniku and Nathan Laundry and Joseph Jay Williams and Anastasia Kuzminykh},
      year={2025},
      eprint={2504.12488},
      archivePrefix={arXiv},
      primaryClass={cs.HC},
      url={https://arxiv.org/abs/2504.12488}, 
}

@article{Rimmershaw1992collabwriting,
    title = {Collaborative writing practices and writing support technologies},
    volume = {21},
    issn = {1573-1952},
    url = {https://doi.org/10.1007/BF00119653},
    doi = {10.1007/BF00119653},
    language = {en},
    number = {1},
    urldate = {2025-07-18},
    journal = {Instructional Science},
    author = {Rimmershaw, Rachel},
    month = jan,
    year = {1992},
    pages = {15--28}
}

@article{Schelble2022thinktogether,
author = {Schelble, Beau G. and Flathmann, Christopher and McNeese, Nathan J. and Freeman, Guo and Mallick, Rohit},
title = {Let's Think Together! Assessing Shared Mental Models, Performance, and Trust in Human-Agent Teams},
year = {2022},
issue_date = {January 2022},
publisher = {Association for Computing Machinery},
address = {New York, NY, USA},
volume = {6},
number = {GROUP},
url = {https://doi.org/10.1145/3492832},
doi = {10.1145/3492832},
journal = {Proc. ACM Hum.-Comput. Interact.},
month = jan,
articleno = {13},
numpages = {29}
}

@inproceedings{Shakeri2021saga,
author = {Shakeri, Hanieh and Neustaedter, Carman and DiPaola, Steve},
title = {SAGA: Collaborative Storytelling with GPT-3},
year = {2021},
isbn = {9781450384797},
publisher = {Association for Computing Machinery},
address = {New York, NY, USA},
url = {https://doi.org/10.1145/3462204.3481771},
doi = {10.1145/3462204.3481771},
booktitle = {Companion Publication of the 2021 Conference on Computer Supported Cooperative Work and Social Computing},
pages = {163–166},
numpages = {4},
location = {Virtual Event, USA},
series = {CSCW '21 Companion}
}

@Inbook{Sharples1993collabwriting,
    author="Sharples, M.
    and Goodlet, J. S.
    and Beck, E. E.
    and Wood, C. C.
    and Easterbrook, S. M.
    and Plowman, L.",
    editor="Sharples, Mike",
    title="Research Issues in the Study of Computer Supported Collaborative Writing",
    bookTitle="Computer Supported Collaborative Writing",
    year="1993",
    publisher="Springer London",
    address="London",
    pages="9--28",
    isbn="978-1-4471-2007-0",
    doi="10.1007/978-1-4471-2007-0_2",
    url="https://doi.org/10.1007/978-1-4471-2007-0_2"
}

@inproceedings{Shaer2024aibrainwriting,
author = {Shaer, Orit and Cooper, Angelora and Mokryn, Osnat and Kun, Andrew L and Ben Shoshan, Hagit},
title = {AI-Augmented Brainwriting: Investigating the use of LLMs in group ideation},
year = {2024},
isbn = {9798400703300},
publisher = {Association for Computing Machinery},
address = {New York, NY, USA},
url = {https://doi.org/10.1145/3613904.3642414},
doi = {10.1145/3613904.3642414},
booktitle = {Proceedings of the 2024 CHI Conference on Human Factors in Computing Systems},
articleno = {1050},
numpages = {17},
location = {Honolulu, HI, USA},
series = {CHI '24}
}

@inproceedings{Shin2023collabIdeationWorkshop,
author = {Shin, Joon Gi and Koch, Janin and Lucero, Andr\'{e}s and Dalsgaard, Peter and Mackay, Wendy E.},
title = {Integrating AI in Human-Human Collaborative Ideation},
year = {2023},
isbn = {9781450394222},
publisher = {Association for Computing Machinery},
address = {New York, NY, USA},
url = {https://doi.org/10.1145/3544549.3573802},
doi = {10.1145/3544549.3573802},
booktitle = {Extended Abstracts of the 2023 CHI Conference on Human Factors in Computing Systems},
articleno = {355},
numpages = {5},
location = {Hamburg, Germany},
series = {CHI EA '23}
}

@inproceedings{Shin2023introbot,
author = {Shin, Donghoon and Kim, Soomin and Shang, Ruoxi and Lee, Joonhwan and Hsieh, Gary},
title = {IntroBot: Exploring the Use of Chatbot-assisted Familiarization in Online Collaborative Groups},
year = {2023},
isbn = {9781450394215},
publisher = {Association for Computing Machinery},
address = {New York, NY, USA},
url = {https://doi.org/10.1145/3544548.3580930},
doi = {10.1145/3544548.3580930},
booktitle = {Proceedings of the 2023 CHI Conference on Human Factors in Computing Systems},
articleno = {613},
numpages = {13},
location = {Hamburg, Germany},
series = {CHI '23}
}

@article{Strobl2019survey,
title = {Digital support for academic writing: A review of technologies and pedagogies},
journal = {Computers \& Education},
volume = {131},
pages = {33-48},
year = {2019},
issn = {0360-1315},
doi = {https://doi.org/10.1016/j.compedu.2018.12.005},
url = {https://www.sciencedirect.com/science/article/pii/S036013151830318X},
author = {Carola Strobl and Emilie Ailhaud and Kalliopi Benetos and Ann Devitt and Otto Kruse and Antje Proske and Christian Rapp}
}

@article{Sun2024metawriter,
author = {Sun, Lu and Tao, Stone and Hu, Junjie and Dow, Steven P.},
title = {MetaWriter: Exploring the Potential and Perils of AI Writing Support in Scientific Peer Review},
year = {2024},
issue_date = {April 2024},
publisher = {Association for Computing Machinery},
address = {New York, NY, USA},
volume = {8},
number = {CSCW1},
url = {https://doi.org/10.1145/3637371},
doi = {10.1145/3637371},
journal = {Proc. ACM Hum.-Comput. Interact.},
month = apr,
articleno = {94},
numpages = {32}
}

@inproceedings{Tankelevitch2024metacog,
author = {Tankelevitch, Lev and Kewenig, Viktor and Simkute, Auste and Scott, Ava Elizabeth and Sarkar, Advait and Sellen, Abigail and Rintel, Sean},
title = {The Metacognitive Demands and Opportunities of Generative AI},
year = {2024},
isbn = {9798400703300},
publisher = {Association for Computing Machinery},
address = {New York, NY, USA},
url = {https://doi.org/10.1145/3613904.3642902},
doi = {10.1145/3613904.3642902},
booktitle = {Proceedings of the 2024 CHI Conference on Human Factors in Computing Systems},
articleno = {680},
numpages = {24},
location = {Honolulu, HI, USA},
series = {CHI '24}
}

@inproceedings{Vaithilingam2024dynavis,
author = {Vaithilingam, Priyan and Glassman, Elena L. and Inala, Jeevana Priya and Wang, Chenglong},
title = {DynaVis: Dynamically Synthesized UI Widgets for Visualization Editing},
year = {2024},
isbn = {9798400703300},
publisher = {Association for Computing Machinery},
address = {New York, NY, USA},
url = {https://doi.org/10.1145/3613904.3642639},
doi = {10.1145/3613904.3642639},
booktitle = {Proceedings of the 2024 CHI Conference on Human Factors in Computing Systems},
articleno = {985},
numpages = {17},
location = {Honolulu, HI, USA},
series = {CHI '24}
}

@article{Venkatraman2024collabstory,
  title={CollabStory: Multi-LLM Collaborative Story Generation and Authorship Analysis},
  author={Venkatraman, Saranya and Tripto, Nafis Irtiza and Lee, Dongwon},
  journal={arXiv preprint arXiv:2406.12665},
  year={2024}
}

@article{Wan2024secondmind,
author = {Wan, Qian and Hu, Siying and Zhang, Yu and Wang, Piaohong and Wen, Bo and Lu, Zhicong},
title = {"It Felt Like Having a Second Mind": Investigating Human-AI Co-creativity in Prewriting with Large Language Models},
year = {2024},
issue_date = {April 2024},
publisher = {Association for Computing Machinery},
address = {New York, NY, USA},
volume = {8},
number = {CSCW1},
url = {https://doi.org/10.1145/3637361},
doi = {10.1145/3637361},
journal = {Proc. ACM Hum.-Comput. Interact.},
month = apr,
articleno = {84},
numpages = {26}
}

@article{Wang2017whyusersdonot,
author = {Wang, Dakuo and Tan, Haodan and Lu, Tun},
title = {Why Users Do Not Want to Write Together When They Are Writing Together: Users' Rationales for Today's Collaborative Writing Practices},
year = {2017},
issue_date = {November 2017},
publisher = {Association for Computing Machinery},
address = {New York, NY, USA},
volume = {1},
number = {CSCW},
url = {https://doi.org/10.1145/3134742},
doi = {10.1145/3134742},
journal = {Proc. ACM Hum.-Comput. Interact.},
month = dec,
articleno = {107},
numpages = {18}
}

@inproceedings{Yuan2022wordcraft,
author = {Yuan, Ann and Coenen, Andy and Reif, Emily and Ippolito, Daphne},
title = {Wordcraft: Story Writing With Large Language Models},
year = {2022},
isbn = {9781450391443},
publisher = {Association for Computing Machinery},
address = {New York, NY, USA},
url = {https://doi.org/10.1145/3490099.3511105},
doi = {10.1145/3490099.3511105},
booktitle = {Proceedings of the 27th International Conference on Intelligent User Interfaces},
pages = {841–852},
numpages = {12},
location = {Helsinki, Finland},
series = {IUI '22}
}

@inproceedings{ZamfirescuPereira2023johnny,
author = {Zamfirescu-Pereira, J.D. and Wong, Richmond Y. and Hartmann, Bjoern and Yang, Qian},
title = {Why Johnny Can’t Prompt: How Non-AI Experts Try (and Fail) to Design LLM Prompts},
year = {2023},
isbn = {9781450394215},
publisher = {Association for Computing Machinery},
address = {New York, NY, USA},
url = {https://doi.org/10.1145/3544548.3581388},
doi = {10.1145/3544548.3581388},
booktitle = {Proceedings of the 2023 CHI Conference on Human Factors in Computing Systems},
articleno = {437},
numpages = {21},
location = {Hamburg, Germany},
series = {CHI '23}
}

@article{Flesch1948readingease,
   author = {Rudolph Flesch},
   doi = {10.1037/H0057532},
   issn = {00219010},
   issue = {3},
   journal = {Journal of Applied Psychology},
   month = {6},
   pages = {221-233},
   pmid = {18867058},
   title = {A new readability yardstick},
   volume = {32},
   year = {1948}
}

@article{Lehmann2025_studyalign,
author = {Lehmann, Florian and Buschek, Daniel},
title = {StudyAlign: A Software System for Conducting Web-Based User Studies with Functional Interactive Prototypes},
year = {2025},
issue_date = {June 2025},
publisher = {Association for Computing Machinery},
address = {New York, NY, USA},
volume = {9},
number = {4},
url = {https://doi.org/10.1145/3733053},
doi = {10.1145/3733053},
journal = {Proc. ACM Hum.-Comput. Interact.},
month = jun,
articleno = {EICS007},
numpages = {26}
}

@manual{yjsdocs,
  title        = {Yjs Documentation},
  author       = {Yjs Contributors},
  year         = {2025},
  url          = {https://docs.yjs.dev/},
  note         = {Accessed: March 25, 2025}
}

@article{sus_eval,
	title        = {An Empirical Evaluation of the System Usability Scale},
	author       = {Aaron Bangor, Philip T. Kortum and Miller, James T.},
	year         = 2008,
	journal      = {International Journal of Human–Computer Interaction},
	publisher    = {Taylor \& Francis},
	volume       = 24,
	number       = 6,
	pages        = {574--594},
	doi          = {10.1080/10447310802205776},
	url          = {https://doi.org/10.1080/10447310802205776},
	eprint       = {https://doi.org/10.1080/10447310802205776}
}

@inproceedings{csi,
	title        = {Creativity factor evaluation: towards a standardized survey metric for creativity support},
	author       = {Carroll, Erin A. and Latulipe, Celine and Fung, Richard and Terry, Michael},
	year         = 2009,
	booktitle    = {Proceedings of the Seventh ACM Conference on Creativity and Cognition},
	location     = {Berkeley, California, USA},
	publisher    = {Association for Computing Machinery},
	address      = {New York, NY, USA},
	series       = {C\&C '09},
	pages        = {127–136},
	doi          = {10.1145/1640233.1640255},
	isbn         = 9781605588650,
	url          = {https://doi.org/10.1145/1640233.1640255},
	numpages     = 10
}

@inproceedings{horvitz1999principles,
author = {Horvitz, Eric},
title = {Principles of mixed-initiative user interfaces},
year = {1999},
isbn = {0201485591},
publisher = {Association for Computing Machinery},
address = {New York, NY, USA},
url = {https://doi.org/10.1145/302979.303030},
doi = {10.1145/302979.303030},
booktitle = {Proceedings of the SIGCHI Conference on Human Factors in Computing Systems},
pages = {159–166},
numpages = {8},
location = {Pittsburgh, Pennsylvania, USA},
series = {CHI '99}
}

@inproceedings{Kabir2024chatgpt,
author = {Kabir, Samia and Udo-Imeh, David N. and Kou, Bonan and Zhang, Tianyi},
title = {Is Stack Overflow Obsolete? An Empirical Study of the Characteristics of ChatGPT Answers to Stack Overflow Questions},
year = {2024},
isbn = {9798400703300},
publisher = {Association for Computing Machinery},
address = {New York, NY, USA},
url = {https://doi.org/10.1145/3613904.3642596},
doi = {10.1145/3613904.3642596},
booktitle = {Proceedings of the 2024 CHI Conference on Human Factors in Computing Systems},
articleno = {935},
numpages = {17},
location = {Honolulu, HI, USA},
series = {CHI '24}
}

@misc{Sun2023evaluating,
      title={Evaluating Large Language Models on Controlled Generation Tasks}, 
      author={Jiao Sun and Yufei Tian and Wangchunshu Zhou and Nan Xu and Qian Hu and Rahul Gupta and John Frederick Wieting and Nanyun Peng and Xuezhe Ma},
      year={2023},
      eprint={2310.14542},
      archivePrefix={arXiv},
      primaryClass={cs.CL},
      url={https://arxiv.org/abs/2310.14542}, 
}

@inproceedings{Prasongpongchai2025hand,
author = {Prasongpongchai, Thanawit and Pataranutaporn, Pat and Lertsutthiwong, Monchai and Maes, Pattie},
title = {Talk to the Hand: an LLM-powered Chatbot with Visual Pointer as Proactive Companion for On-Screen Tasks},
year = {2025},
isbn = {9798400713941},
publisher = {Association for Computing Machinery},
address = {New York, NY, USA},
url = {https://doi.org/10.1145/3706598.3715579},
doi = {10.1145/3706598.3715579},
booktitle = {Proceedings of the 2025 CHI Conference on Human Factors in Computing Systems},
articleno = {637},
numpages = {16},
location = {
},
series = {CHI '25}
}

@inproceedings{gero_sparks_2022,
	address = {Virtual Event Australia},
	title = {Sparks: {Inspiration} for {Science} {Writing} using {Language} {Models}},
	isbn = {978-1-4503-9358-4},
	shorttitle = {Sparks},
	url = {https://dl.acm.org/doi/10.1145/3532106.3533533},
	doi = {10.1145/3532106.3533533},
	language = {en},
	urldate = {2022-07-19},
	booktitle = {Designing {Interactive} {Systems} {Conference}},
	publisher = {ACM},
	author = {Gero, Katy Ilonka and Liu, Vivian and Chilton, Lydia},
	month = jun,
	year = {2022},
	pages = {1002--1019},
}

@inproceedings{dang_choice_2023,
	address = {Hamburg Germany},
	title = {Choice {Over} {Control}: {How} {Users} {Write} with {Large} {Language} {Models} using {Diegetic} and {Non}-{Diegetic} {Prompting}},
	isbn = {978-1-4503-9421-5},
	shorttitle = {Choice {Over} {Control}},
	url = {https://dl.acm.org/doi/10.1145/3544548.3580969},
	doi = {10.1145/3544548.3580969},
	language = {en},
	urldate = {2025-02-06},
	booktitle = {Proceedings of the 2023 {CHI} {Conference} on {Human} {Factors} in {Computing} {Systems}},
	publisher = {ACM},
	author = {Dang, Hai and Goller, Sven and Lehmann, Florian and Buschek, Daniel},
	month = apr,
	year = {2023},
	pages = {1--17},
}

@inproceedings{Siddiqui2025scriptshift,
author = {Siddiqui, Momin N and Pea, Roy D and Subramonyam, Hari},
title = {Script\&Shift: A Layered Interface Paradigm for Integrating Content Development and Rhetorical Strategy with LLM Writing Assistants},
year = {2025},
isbn = {9798400713941},
publisher = {Association for Computing Machinery},
address = {New York, NY, USA},
url = {https://doi.org/10.1145/3706598.3714119},
doi = {10.1145/3706598.3714119},
booktitle = {Proceedings of the 2025 CHI Conference on Human Factors in Computing Systems},
articleno = {532},
numpages = {19},
location = {
},
series = {CHI '25}
}

@article{Yeung2018readability,
	title = {Readability of the 100 {Most}-{Cited} {Neuroimaging} {Papers} {Assessed} by {Common} {Readability} {Formulae}},
	volume = {12},
	issn = {1662-5161},
	url = {https://pmc.ncbi.nlm.nih.gov/articles/PMC6104455/},
	doi = {10.3389/fnhum.2018.00308},
	urldate = {2025-11-08},
	journal = {Frontiers in Human Neuroscience},
	author = {Yeung, Andy W. K. and Goto, Tazuko K. and Leung, W. Keung},
	month = aug,
	year = {2018},
	pmid = {30158861},
	pmcid = {PMC6104455},
	pages = {308}
}

@article{PlavenSigray2017readability,
	title = {The readability of scientific texts is decreasing over time},
	volume = {6},
	issn = {2050-084X},
	url = {https://pmc.ncbi.nlm.nih.gov/articles/PMC5584989/},
	doi = {10.7554/eLife.27725},
	urldate = {2025-11-08},
	journal = {eLife},
	author = {Plavén-Sigray, Pontus and Matheson, Granville James and Schiffler, Björn Christian and Thompson, William Hedley},
	pmid = {28873054},
	pmcid = {PMC5584989},
	pages = {e27725},
    year = {2017}
}

@inproceedings{Feng2025canvil,
author = {Feng, K. J. Kevin and Liao, Q. Vera and Xiao, Ziang and Wortman Vaughan, Jennifer and Zhang, Amy X. and McDonald, David W.},
title = {Canvil: Designerly Adaptation for LLM-Powered User Experiences},
year = {2025},
isbn = {9798400713941},
publisher = {Association for Computing Machinery},
address = {New York, NY, USA},
url = {https://doi.org/10.1145/3706598.3713139},
doi = {10.1145/3706598.3713139},
booktitle = {Proceedings of the 2025 CHI Conference on Human Factors in Computing Systems},
articleno = {932},
numpages = {22},
location = {
},
series = {CHI '25}
}

@inproceedings{Hutchinson2003techprobe,
author = {Hutchinson, Hilary and Mackay, Wendy and Westerlund, Bo and Bederson, Benjamin B. and Druin, Allison and Plaisant, Catherine and Beaudouin-Lafon, Michel and Conversy, St\'{e}phane and Evans, Helen and Hansen, Heiko and Roussel, Nicolas and Eiderb\"{a}ck, Bj\"{o}rn},
title = {Technology probes: inspiring design for and with families},
year = {2003},
isbn = {1581136307},
publisher = {Association for Computing Machinery},
address = {New York, NY, USA},
url = {https://doi.org/10.1145/642611.642616},
doi = {10.1145/642611.642616},
booktitle = {Proceedings of the SIGCHI Conference on Human Factors in Computing Systems},
pages = {17–24},
numpages = {8},
location = {Ft. Lauderdale, Florida, USA},
series = {CHI '03}
}

@misc{ifttt2025,
	title = {{IFTTT} - {Automate} business \& home},
	url = {https://ifttt.com/},
	language = {en},
	urldate = {2025-11-10},
	journal = {IFTTT},
	author = {IFTTT},
    year = {2025},
}

@misc{Claude2025promptlibrary,
	title = {Claude Prompt Library},
	url = {https://docs.claude.com/en/resources/prompt-library/library},
	language = {en},
	urldate = {2025-11-10},
	journal = {Claude Docs},
    year = {2025},
}
